\documentclass[times,final,authoryear,a4paper]{article}
\usepackage[pdftex]{graphicx}
\usepackage{framed,multirow}
\usepackage[utf8]{inputenc}
\usepackage[utf8]{inputenc}

\usepackage{amssymb}
\usepackage{latexsym}

\usepackage{url}
\usepackage{xcolor}
\definecolor{newcolor}{rgb}{.8,.349,.1}

\usepackage[citebordercolor=white]{hyperref}

\DeclareUnicodeCharacter{200B}{{\hskip 0pt}}
\DeclareUnicodeCharacter{0308}{{\hskip 0pt}}

\newcommand{\fig}[1]{Fig.\,\ref{fig:#1}}
\newcommand{\tab}[1]{Tab.\,\ref{tab:#1}}
\begin{document}

\title{HERO (High Energy Ray Observatory) optimization and\\ current status}%
\author{Alexander Kurganov${}^1$
\thanks{Corresponding author: e-mail: me@sx107.ru},
Dmitry Karmanov${}^1$,
Alexander Panov${}^1$,
Dmitry Podorozhny${}^1$,\\
Leonid Tkachev${}^2$,
Andrey Turundaevsky${}^1$}

\date{}

\maketitle

\begin{center}
${}^1$Lomonosov Moscow State University Skobeltsyn Institute of Nuclear Physics (MSU SINP),Leninskie gory 1(2), GSP-1, Moscow, 119991, Russia.\\
${}^1$Joint Institute for Nuclear Research, Joliot-Curie 6,
Dubna, Moscow Region, 141980, Russia.
\end{center}

\sloppy
\begin{abstract}
The High-Energy Ray Observatory (HERO) is a space experiment based on a heavy ionization calorimeter for direct study of cosmic rays. The effective geometrical factor of the apparatus varies from 12 to 60 m$^2$sr for protons depending on the weight of the calorimeter from 10 to 70 tons. During the exposure for $\sim$5 years this mission will make it possible to measure energy spectra of all abundant cosmic ray nuclei in the knee region ($\sim$3 PeV) with individual resolution of charges with energy resolution better than 30\% and provide useful information to solve the puzzle of the cosmic ray knee origin. HERO mission will make it also possible to measure energy spectra of cosmic rays nuclei for energies 1-1000 TeV with very high precision and energy resolution (up to 3\% for calorimeter 70 tons) and study the fine structure of the spectra. The planned experiment launch is no earlier than 2029.
\end{abstract}

\twocolumn
%% main text
\section{Introduction}

Galactic cosmic rays (GCR) refer to flows of ultrarelativistic charged particles that fill interstellar space. Cosmic rays are represented by hadronic and lepton components. The hadronic (otherwise -- nuclear) component of the GCR consists of chemical elements nuclei from protons (hydrogen nuclei) to nuclei much heavier than iron. The lepton component consists of electrons and positrons. Although the hadronic component is highly dominant, the presence of a lepton component is also very important for understanding the nature of cosmic rays, since it can contain important information about the nearest sources of cosmic rays and about some exotic objects in space such as dark matter or primordial black holes. There may also be some hypothetical exotic objects among relativistic cosmic particles, such as stranglets. Cosmic rays in the Galaxy correspond to approximately the same average energy distribution density (on the scale of one $\textrm{eV/cm}^3$) as the energy density of the magnetic field, energy density of the radiation field (light, etc.) and the average density of the kinetic energy of ordinary matter. In this sense, cosmic rays are one of the main components of outer space, and the physics of cosmic rays requires deep study to understand the world in which we live - our Galaxy. This explains the natural general scientific interest in cosmic rays.

The most likely source of the major part of cosmic rays are supernovae explosions in our Galaxy \cite{GINZBURG1964}. The supernovae astrophysics is of great interest for many reasons. Supernovae are the source of heavy chemical elements that make up terrestrial planets and ourselves. Supernova explosions represent one of the most important phases in the evolution of large stars, as a result of which relativistic astrophysical objects such as neutron stars and black holes can be formed -- these objects and their origin routes are of great interest, since they are associated with problems that lead to the limits of current understanding of the essence of space, time and matter. Explosions of nearby supernovae in the past could have a significant impact on the course of biological evolution on Earth and may have such an impact in the future, which is important. In this sense, from a purely practical point of view, it is necessary to understand well what types of supernovae exist, which consequences may be associated with their explosions, and what can we expect from our closest stellar environment. The closest sources of cosmic rays are of particular interest from this point of view. Since supernova explosions are the main source of cosmic rays, it is cosmic rays that can and do provide a large amount of information about the physics of supernovae. Recently, interest has shifted to the study of fine structure of the cosmic ray nuclei energy spectra and the chemical composition of cosmic rays. These spectral features encode subtle details of supernovae explosions physics and the acceleration of cosmic rays, such as, for example, the presence of different types of supernovae with different thresholds of acceleration energies and different composition of sources.

The propagation of cosmic rays in the Galaxy is essentially determined by the properties of the interstellar medium and the structure of the Galaxy. Therefore, a number of cosmic ray spectra features make it possible to obtain data on the magnitude and structure of interstellar magnetic fields, the density of interstellar matter, the presence and extent of the Galactic magnetic halo, matter fluxes, etc.

Finally, the cosmic rays scientific problems are closely related to numerous areas of research on cosmic gamma radiation. The main source of gamma rays are the regions where ordinary charged cosmic rays are accelerated, but there may be other important sources of cosmic gamma rays, such as, for example, decay or annihilation of dark matter particles. The study of cosmic gamma radiation therefore not only sheds additional light on the nature of sources and the physics of cosmic rays themselves, but can also provide information on the nature of dark matter. The main difference between gamma quanta and charged cosmic rays is that gamma quanta propagate almost in a straight line (up to deflection in gravitational fields due to the curvature of space), so the direction of their arrival very accurately indicates their source. The measurement of cosmic gamma spectrum with the highest possible energy resolution is of particular importance, since a narrow peak in such a spectrum can be associated with gamma quanta from the dark matter particles annihilation and provide immediate information about their mass. If the direction of arrival of quanta belonging to this peak will, moreover, mainly point to the center of the Galaxy, where a very powerful cusp (compact concentration) in the dark matter density distribution is expected, then the probability of interpreting such a peak to be associated with the annihilation of dark matter will be very significant.

It has been known since the 1950s that one of the main features of the energy spectrum of cosmic rays is a kink (knee) at energies between $10^{15}$ and $10^{16}$ eV \cite{KULIKOV-HRISTIANSEN1958} against the background that as a whole is approximately a power-law spectrum. However, in the 2000s, it became clear that even at lower energies, the cosmic ray spectra have many features that violate the simple universal power-law behavior. The first such feature to be discovered was the difference in the slopes of the spectra of protons and helium at energies from about 100 GeV to 10 TeV, discovered for the first time in the ATIC experiment \cite{ATIC-2004-ZATSEPIN-IzvRan}. Later the effect was confirmed in the experiments CREAM \cite{CREAM2009B,CREAM2011-ApJ-PHe-I,CR-CREAM2017-ApJ-pHe}, PAMELA \cite{CR-PAMELA-2011-p-He-Mag}, AMS-02 \cite{AMS-02-2015-PRL-p,AMS-02-2015-PRL-He} and others. In a later ATIC article \cite{ATIC-2017-ApJ}, a systematic change in the slope of the spectrum from helium to iron with increasing nuclear charge was found in the magnetic rigidity range from 50 to 1350 GV. In the article of the ATIC collaboration \cite{ATIC-2007-PANOV-IzvRAN} it was noted for the first time that the spectra of protons and helium at energies from 50 GeV to about 10 TeV have a shape that significantly differs from the power-law form in the form of an upturn at energies between 100 GeV and 1 TeV. A similar upturn was found in the spectra of heavier abundant nuclei \cite{ATIC-2009-PANOV-IzvRAN-ENG}. These definite flattenings were later confirmed by the results of CREAM \cite{CR-CREAM2010A}, PAMELA \cite{CR-PAMELA-2011-p-He-Mag} and AMS-02 \cite{AMS-02-2015-PRL-p,AMS-02-2015-PRL-He}.  There was a strong indication of the existence of breaks in the  protons and helium spectra near the magnetic ridgidity of 10 TV in an article from the CREAM experiment \cite{CR-CREAM2017-ApJ-pHe}. The existence of these breaks was confirmed at a level of about four standard deviations in the NUCLEON experiment \cite{NUCLEON-KNEE-2018,NUCLEON-ABDNT-2019,NUCLEON-REW-2021}. In the same articles of NUCLEON, it was shown that a similar knee near the same rigidity 10 TV (at the same significance level) exists not only in the spectra of protons and helium, but also in the total rigidity spectrum of all heavy nuclei from carbon to iron. That is, a knee near 10 TV is of a universal nature (similar fact is still not tested and not known about the main knee of cosmic rays near 3 PeV energy per particle). The existence of this ``small'' knee in the spectra of protons and helium was confirmed in the DAMPE experiment \cite{DAMPE-p-2019,DAMPE-He-2021}. The existence of a small knee in the spectrum of heavy nuclei, which is observed in the NUCLEON experiment, has not yet been tested by other experiments. Thus, various experiments qualitatively consistently confirm the presence of a fine structure of the spectra in the energy range from hundreds of GeV to about 100 TeV; however, the quantitative details of the behavior of these structures in different experiments are still different. Therefore, more accurate direct observations of cosmic rays are relevant in this energy range.

In the region of the large knee of cosmic rays (around 3 PeV), the main data is obtained in the extensive air shower (EAS) experiments (see \cite{TUNKA-2019} for an overview of the latest results). These experiments currently provide high statistical significance and reliability in measuring the energy spectrum of all particles. However, they provide only very poor averaged information on the chemical composition of nuclei, without giving element-by-element charge resolution. Despite the fact that recently a very good matching of data from the direct and EAS measurements has been achieved for the total spectrum of all particles \cite{NUCLEON-REW-2021}, the element-by-element structure of the 3 PeV knee is still unknown. This greatly complicates the interpretation of this most important feature of the cosmic ray spectrum.

Thus, two very important tasks can be distinguished in the physics of high-energy cosmic rays: measuring the precision spectra of cosmic rays at energies below the Kulikov-Hristiansen 3 PeV knee, and measuring spectra with elemental charge identification in the region of the 3 PeV knee. These two problems define the main scientific problems to be solved by the HERO (High-Energy Ray Observatory) experiment.

The HERO space observatory is based on the use of an ultra-heavy image calorimeter to measure the energy of particles, reconstruct their trajectories and separate the electromagnetic and hadronic CR components. The configurations with calorimeter weights of 10, 30 and 70 tons are currently being considered in designing the instrument. Well-proven techniques based on silicon detector arrays will be used to determine the particle charge, quite similar to those used in the NUCLEON experiment (see \cite{NUCLEON-DEZ-2007A,NUCLEON-DEZ-2007B,NUCLEON-DEZ-2015}), in which a charge resolution of 0.2 charge units was achieved. The charge resolution of HERO at energies of cosmic rays up to several hundreds of TeV should be only better, since the single active pad size is planned to be halved. The study of exact charge distributions at multi-PeV energies requires a very long computational time; it is currently in progress and results will be published later, but preliminary estimates (discussed later in this article, see \hyperref[sec:backcur]{subsection 2.3}) show that backscattered particles will not interfere with the identification of even cosmic ray protons, not to mention more heavy nuclei. It should also be noted that the problem of back currents at ultrahigh energies is not critical, since the back current density can be controlled by the distance from the charge detector to the calorimeter, and it is only necessary to choose the optimal distance.

The HERO experiment will also be able to solve some other actual problems of cosmic ray physics in addition to the two main tasks mentioned above. These include measuring the spectra of cosmic ray leptons as well as spectra of gamma quanta with an ultra-high energy resolution on a scale of 1\%, which is important for clarifying the nature of leptons energy spectrum features and search for the dark matter annihilation line in gamma spectra; measurement of the CR anisotropy in direct measurements with a previously unattainable statistical accuracy; study of the secondary CR nuclei spectral features to refine the propagation models. Since the device will allow registering particle charges up to $Z \sim 100$ and isolating particles with sharp anomalies in the mass-to-charge ratio, the experimental data from HERO will allow either to detect exotic particles such as stranglets, or to give new restrictions on their abundance, which will be very difficult to exceed in any experiments in the foreseeable future. Some of these tasks will be discussed below. Without much exaggeration, we can say that the HERO experiment should close the main questions of the classical physics of the nuclear component of the CR, which can potentially be solved in direct extra-atmospheric measurements, as well as provide a lot of unique information for new physics related to such exotic objects as stranglets or dark matter.

None of the modern space experiments in cosmic ray physics are comparable in their capabilities to the HERO observatory. Among the promising instruments, HERO can be compared to the HERD observatory, which is scheduled to launch near 2030 \cite{HERD-2019}. However, even in its minimal configuration with a calorimeter weight of 10 tons, the HERO observatory's geometric factor will surpass the one of HERD by more than 5 times, so even HERD can hardly be considered as a competitor to HERO. No other comparable instruments are planned.

\section{Optimization of the design of the observatory}

\begin{figure}[ht!]
 \begin{center}
 \includegraphics[width=0.45\textwidth]{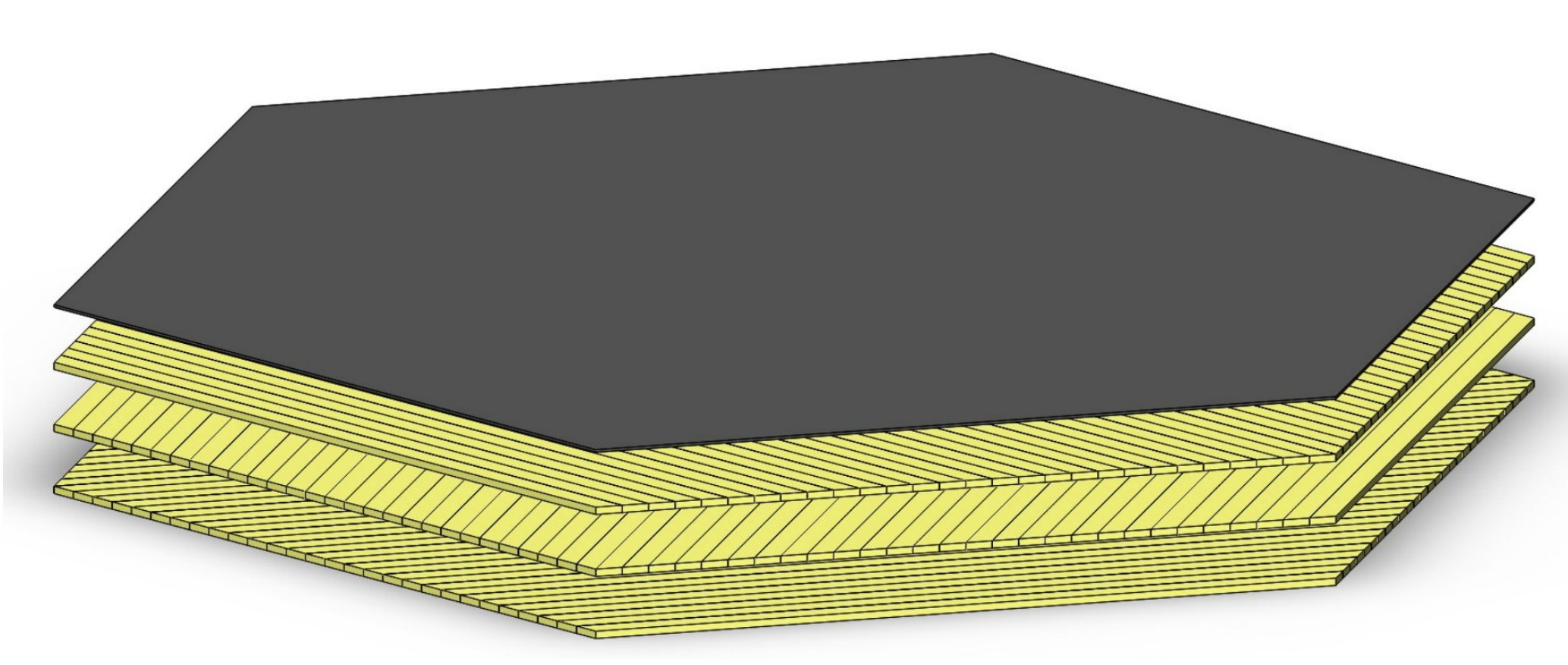}
 \caption{One calorimeter layer for a configuration of three scintillator planes rotated by $60^\mathrm{o}$ relative to each other. \label{fig:Plane}}
 \end{center}
\end{figure}

\begin{figure*}[ht!]
 \begin{center}
 \includegraphics[width=0.8\textwidth]{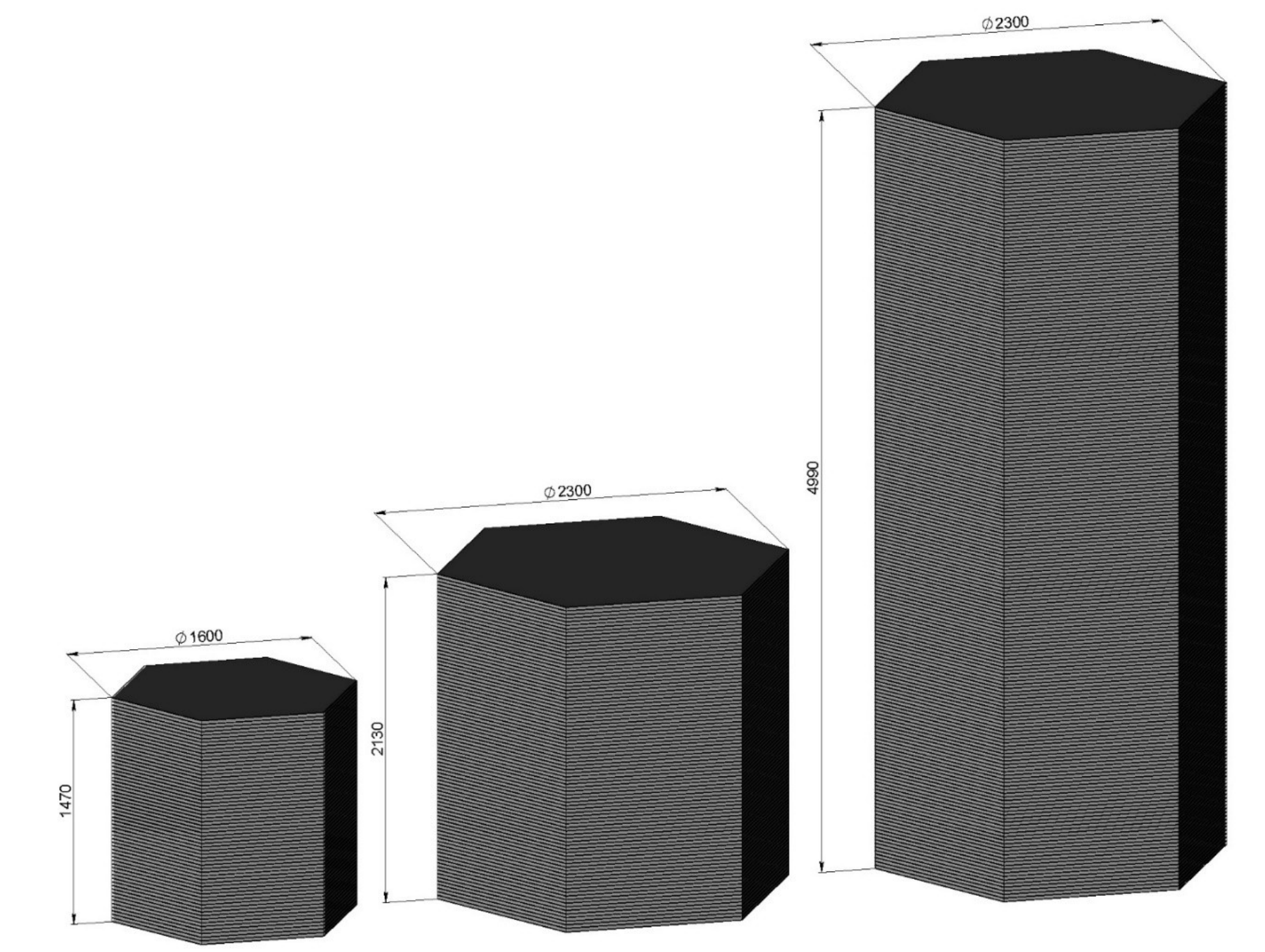}
 \caption{Optimized dimensions and appearance of calorimeters with weights of 10, 30 and 70 tons. All units in millimeters. \label{fig:AllWeights}}
 \end{center}
\end{figure*}

The HERO observatory is based on an image scintillation tungsten ionization calorimeter with a layered structure. During the discussion of the HERO project, the following circumstance was revealed. The previously investigated structure of the calorimeter, represented by a lattice of hexagonal cells, \cite{HERO-2017} does not allow the energy resolution for electrons and gamma quanta to be much better than 10\%. Such a resolution can be considered satisfactory for studying the electrons spectrum, but it is not at all enough for studying the spectrum of gamma quanta. Considering the current estimation of the gamma-ray spectrum from the central region of the Galaxy ($dN/dE = \Phi_1 (E/\mathrm{TeV})^{-{\Gamma_1}}$; $\Phi_1 = (1.92 \pm 0.08_{\mathrm{stat}} \pm 0.28_{\mathrm{syst}})\times 10^{-12} \mathrm{TeV}^{-1}\mathrm{cm}^{-2} \mathrm{s}^{-1}$; $\Gamma_1 = 2.32 \pm 0.05_{\mathrm{stat}} \pm 0.11_{\mathrm{syst}}$: \cite{HESS-2016}), for the apparatus acceptance area of $\sim 10\,\mathrm{m}^2$ (70 tons configuration, see \fig{AllWeights}), ten years of exposition the estimated number of gamma quants will be 9500 for $E>100$\,GeV, 230 for $E>500$\,GeV, 40 for $E>1$\,TeV.  The unique ability to measure the spectrum of gamma quanta from the central regions of the Galaxy up to energies on a scale of TeV, which is provided by the high geometric factor of the device, is not accompanied by the possibility of detecting monoenergetic gamma lines in this spectrum due to too low energy resolution, which significantly devalues the spectrum, which could have been obtained. Therefore, it was decided to return from a cellular structure to a more conventional pure layered one, where layers of a heavy absorber alternate with layers of a scintillator. Each layer of the calorimeter now consists of one plane of tungsten with a thickness of about one radiation unit and two or three planes of a scintillator (polystyrene) with a total thickness of about 20 mm (see details below). Each plane of the scintillator is composed of scintillation strips with different strip orientation in different planes. In the case of two layers, the strips are oriented mutually perpendicular, in the case of three layers, the strips are rotated 60$^\mathrm{o}$ in each subsequent layer (see \fig{Plane}). The horizontal section of the calorimeter has the shape of a regular hexagon, as it was already investigated at earlier stages of the project \cite{HERO-2019}. The appearance of the calorimeters of different weights is shown on \fig{AllWeights}. Simulation of a new device configuration with a layered calorimeter structure using the FLUKA package \cite{FLUKA2014} showed that with a calorimeter mass of 10 tons for any reasonable device configurations, the energy resolution for electrons and gamma quanta will be very high. In accordance with the calculated instrumental functions for gamma quanta with energies of 100 GeV, 1000 GeV, 10000 GeV, they correspond to a resolution of no worse than 0.96\%, 0.31\%, 0.096\%, respectively. Here, only physical fluctuations of the signal are taken into account, but hardware noises are not yet considered. Obviously, for energies above 100 GeV, the energy resolution for gamma quanta and electrons will be determined mainly by the electronics noise. Note that in this case, the effective area and geometric factor of HERO will significantly exceed those of the best modern instruments. For example, the geometric factor of the Fermi Large Area Telescope for electrons with energies of 100 GeV to 1 TeV varies from 3 m$^2$sr to 1 m$^2$sr respectively \cite{FERMI-TINIVELLA-2016}, while for the HERO observatory, depending on the configuration, it will range from $\sim$10 m$^2$sr to $\sim$70 m$^2$sr for all reasonable energies, see section \ref{SizeWeight}.

The number of scintillator planes in each layer of the calorimeter, the width of the scintillator strip and the outer dimensions of the calorimeter are the values optimized for calorimeters of each given weight. 

\subsection{Calorimeter Size Optimization: Geometric Factor and Energy Resolution}
\label{SizeWeight}

Since the calorimeter, generally speaking, has a large thickness in a sense that its lateral surface is large, in order to preserve a large geometric factor, one cannot confine themselves only to events arriving at the calorimeter from above or below. Events arriving through the lateral faces must also be accepted. However, not all such events can be reasonably handled. If, for example, an event impinges almost exactly parallel to the scintillator plane, its cascade curve then cannot be reconstructed. The criterion for events selection was as follows. First, it was required that the length of the particle (proton) trajectory crossing the calorimeter was at least 1.5 nuclear units of length, and secondly, it was required that the trajectory crossed at least 8 layers of scintillators, so that the cascade curve could be reconstructed with required accuracy. Geometry optimization (with the accompanying calculation of the geometric factor) was carried out for protons. Protons are the most difficult objects to register among all other heavier nuclei, since they have the longest nuclear interaction length, therefore, they require a larger calorimeter thickness for a satisfactory measurement of the cascade longitudinal profile. For heavier nuclei, the registration conditions will automatically be better than for protons.  To find the optimal geometry, it was assumed that the observatory's orbit altitude is 500 km, and the Earth's shadow was taken into account in calculating the geometric factor. As already mentioned, the calorimeter is a hexahedral prism, one of the lateral faces of which was turned towards the Earth. The external proton flux was assumed to be isotropic; optimization of the calorimeter was carried out for 2 TeV protons. For some calorimeter configurations, it has been verified that the simulation results are weakly dependent on energy up to energies of 2 PeV.

The carrying capacity of the carrier rocket that will launch the HERO observatory into orbit has not yet been precisely determined, so the permissible weigh of the calorimeter can vary from 10 to 70 tons. The most important characteristics of the observatory, such as the geometric factor, resolution in determining the energy spectra of cosmic ray nuclei, and the optimal shape of the calorimeter, radically depend on the calorimeter weight. This section examines a range of problems related to the dependence of the characteristics of the device on the weight of the calorimeter.

Extensive simulations of various calorimeter shapes were carried out using FLUKA system, to find the optimal configuration for calorimeter weights of 10, 30 and 70 tons. For the choice of the shape of the calorimeter, the most important factor is the optimal combination of geometric factor and energy resolution that can be obtained for a given weight.

This simulation was based on a simplified structure of the calorimeter consisting of alternating layers of tungsten with thickness of one radiation unit (about 3 mm) and a plastic scintillator with thickness of 20 mm without taking into account the strip structure of scintillators, which is not important for this particular problem. In the case of isotropic incidence of particles on a complex-shaped calorimeter, different situations arise in the sense of intersection of particle trajectory with the calorimeter volume. The most important factor here is the length of the part of trajectory that intersects with the calorimeter. It is well known that the energy resolution of an ionization calorimeter at normal incidence of particles at its entrance depends on the thickness of the calorimeter: the thicker the calorimeter, the greater part of the particle energy is absorbed in it, the smaller the relative fluctuations of the absorbed energy and the higher the resolution. When considering an entire set of particles trajectories crossing a complex-shaped calorimeter when an isotropic flux is incident on it, for each length of a part of the trajectory crossing the calorimeter, a situation arises that is equivalent to using a calorimeter of the same thickness with normal incidence of particles. That is, different particle trajectories will be characterized with different energy resolutions, depending on the length of the part of the particle trajectory passing through the calorimeter volume. There won't be any single universal instrument energy resolution for all trajectories, but there will be a probability distribution of the resolutions that can be computed by Monte Carlo simulations. This probability distribution is of primary interest in our problem. Obviously, this probability distribution will radically depend on the shape and mass of the calorimeter.

Before proceeding with the calculation of the instrument resolution distributions, a relationship must be established between the length of the trajectory segment passing through the calorimeter and resolution typical for this length. Or, equivalently, a relationship must be established between the calorimeter thickness and the energy resolution at normal incidence of particles on the calorimeter.

To solve this problem, normal incidence of 2 TeV protons on a calorimeter of the design described above, consisting of alternating layers of tungsten and scintillator, was simulated. Protons were chosen because, as it is clear from above, the energy resolution for protons is the worst among all other nuclei. For all heavier nuclei, the energy resolution at all energies will be better than for protons. In this problem, the calorimeter thickness or trajectory lengths are most conveniently measured in nuclear proton free paths. We investigated calorimeters with a thickness of 1.50 to 15.21 nuclear proton path lengths. A trigger condition was also simulated: at least 500 MIP of energy should be deposited in the calorimeter scintillators at the first nuclear path length (MIP refers to a scintillator 20 mm thick). It is expected that a similar trigger will be implemented in a real HERO experiment. This excludes events with a very late nuclear interaction of a proton with the substance of the calorimeter, or even those that slipped through the calorimeter without nuclear interaction at all and did not generate an electromagnetic-nuclear cascade.

\begin{figure*}[ht!]
 \begin{center}
 \includegraphics[width=0.8\textwidth]{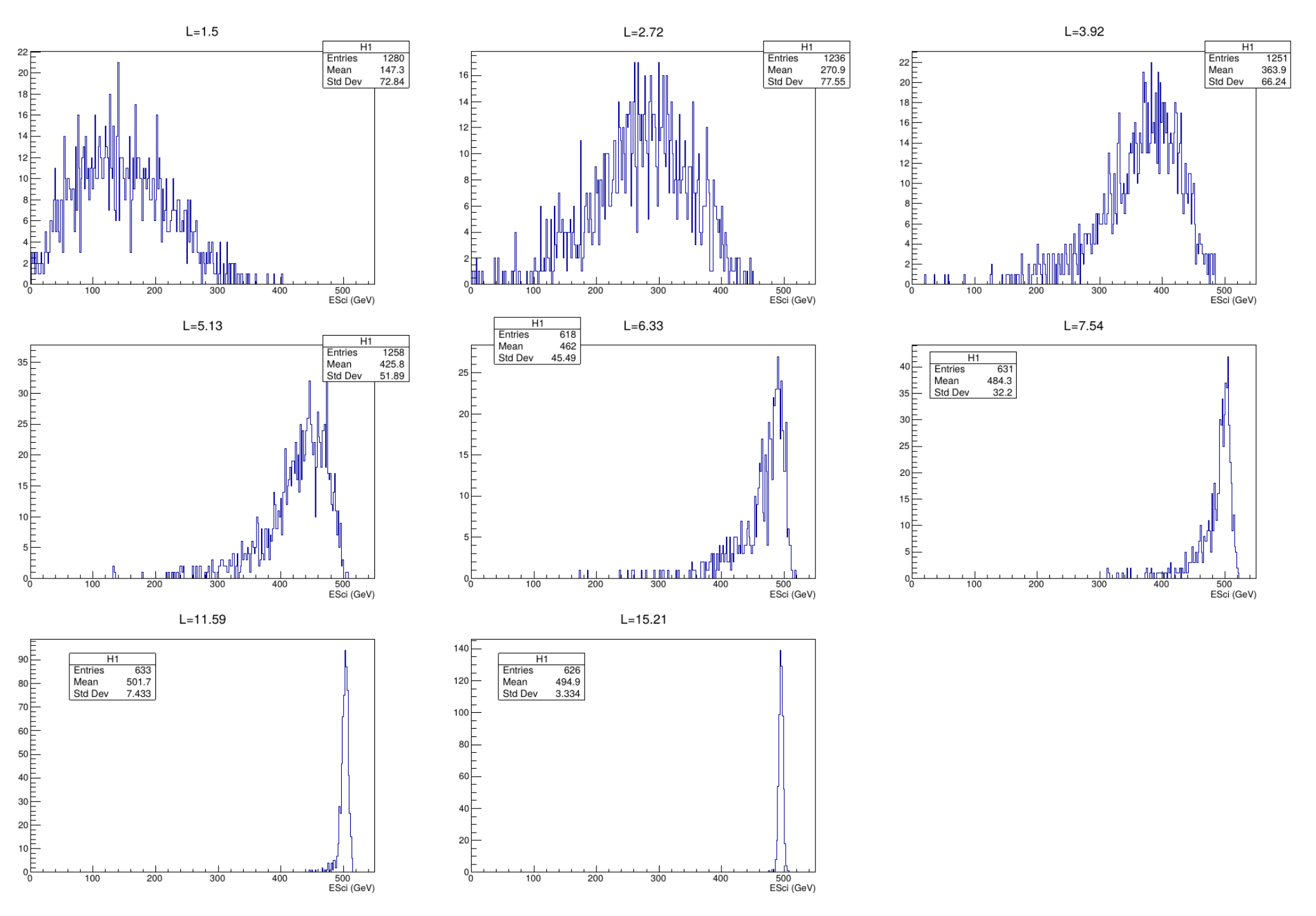}
 \caption{Energy deposition in the scintillators of the calorimeter at a normal incidence of 2 TeV protons on calorimeters of different thicknesses (the thickness measured in nuclear proton path lengths is indicated in the title of each histogram).\label{fig:Res1}}
 \end{center}
\end{figure*}

\begin{figure}
 \begin{center}
 \includegraphics[width=0.45\textwidth]{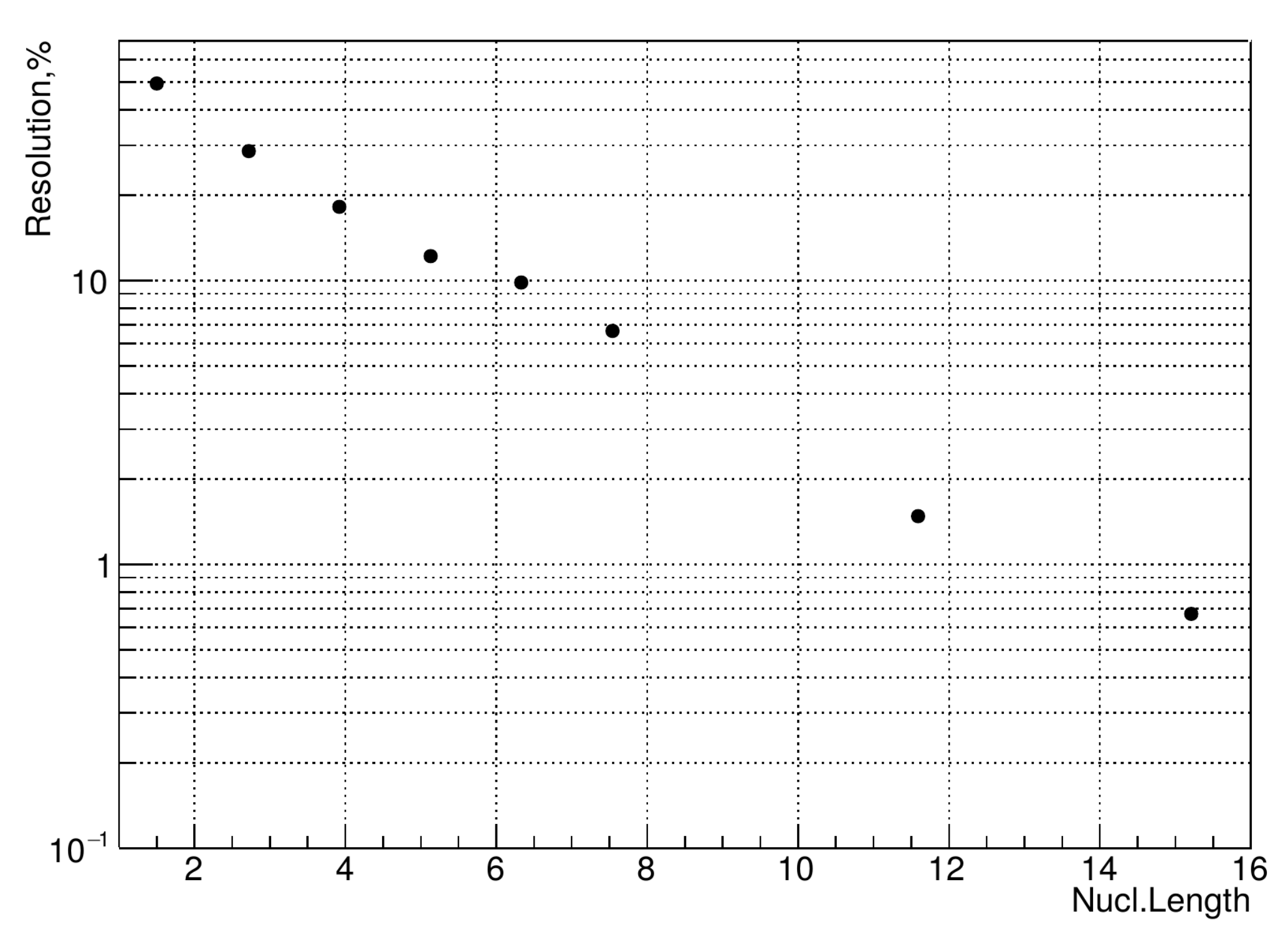}
 \caption{Dependence of the energy resolution of the calorimeter on its thickness (in units of nuclear proton path length) at normal incidence of 2 TeV protons.\label{fig:Res2}}
 \end{center}
\end{figure}

\fig{Res1} shows the obtained distributions of energy deposition in the scintillators for calorimeters of different thicknesses (the thickness in nuclear proton paths is indicated in the title of each histogram). Note that at energies above 1 TeV, the energy resolution of the calorimeter only weakly depends on the particle energy. The relative energy resolution is then calculated as a ratio of the of the energy distribution standard deviation to its mean value. From the histograms shown on \fig{Res1} the dependence of the energy resolution on the calorimeter thickness at normal incidence is obtained -- or, equivalently, the dependence of the resolution on the length of the part of the proton trajectory passing through the calorimeter body with the isotropic incidence of the particle flux onto the calorimeter of arbitrary shape. This dependency is shown on \fig{Res2}. The dependence of the energy resolution on the length of the trajectory in the calorimeter, shown in \fig{Res2}, was interpolated using a cubic spline and, in this form, was used for calculating the energy resolution distributions.

Let us now consider the results of modeling the isotropic incidence of particles on the calorimeter of the HERO observatory and the corresponding optimization of the shape of the calorimeter. The calorimeter shape is based on a regular hexagonal straight prism, but the height and diameter of the circumscribed circle can be different. It was assumed in this optimization and geometry factors calculation that in the Earth's orbit with an altitude of 500 km, the prism is always located parallel to one of its lateral faces of the Earth's surface.

\begin{figure*}[ht!]
 \begin{center}
 \includegraphics[width=0.8\textwidth]{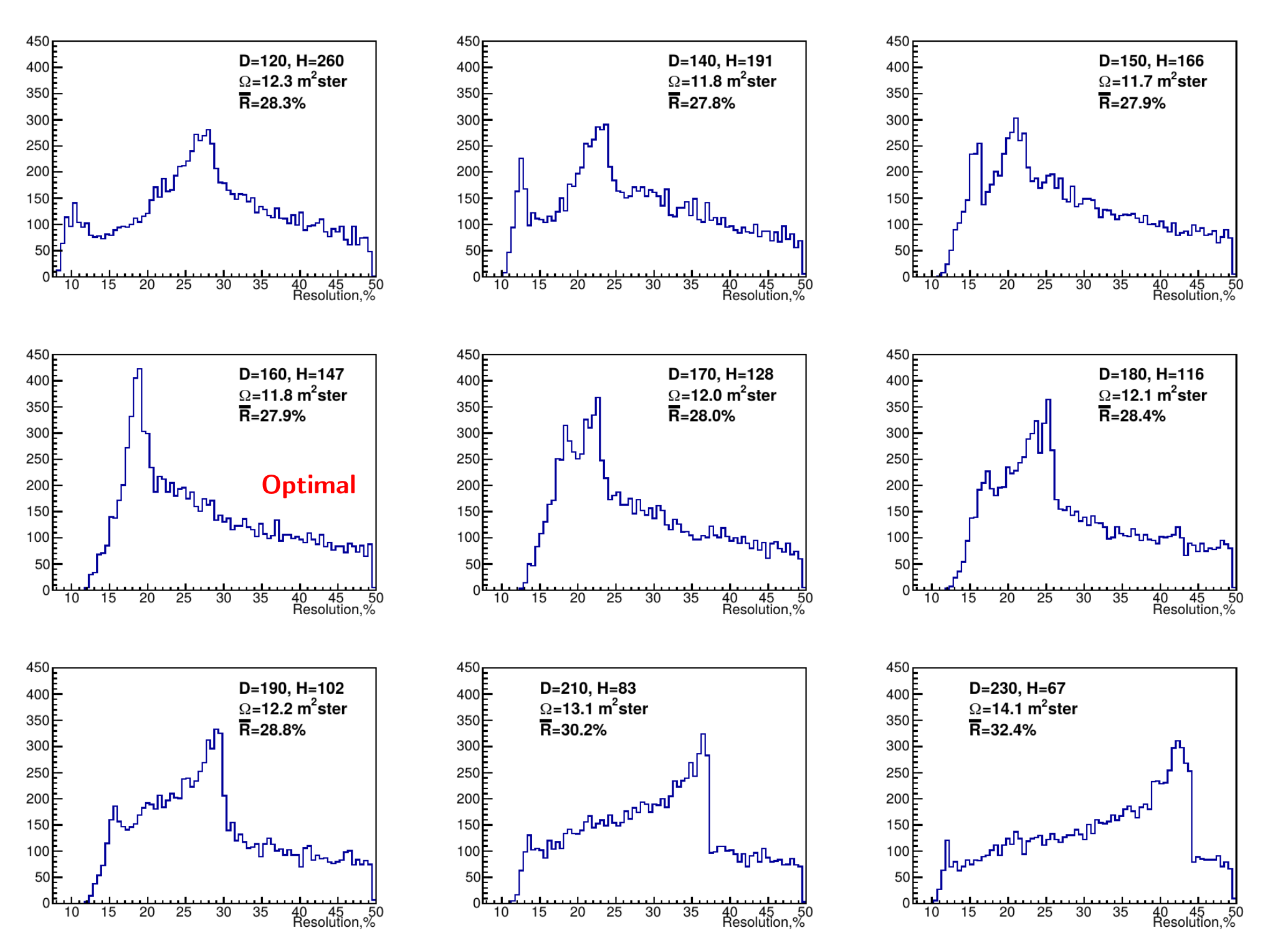}
 \caption{Energy resolution distributions for a calorimeter 10 tons and various shapes of the calorimeter. The diameter of the circumscribed circle of the prism $D$ and the height of the calorimeter $H$ are expressed in centimeters. The graphs also show the effective geometric factor (for protons that have passed all selection conditions, including the trigger condition) and the energy resolution averaged over the distribution. This plot was publeshed first in \cite{HERO-2019} \label{fig:Distrib10}}
 \end{center}
\end{figure*}

\begin{figure*}[ht!]
 \begin{center}
 \includegraphics[width=0.8\textwidth]{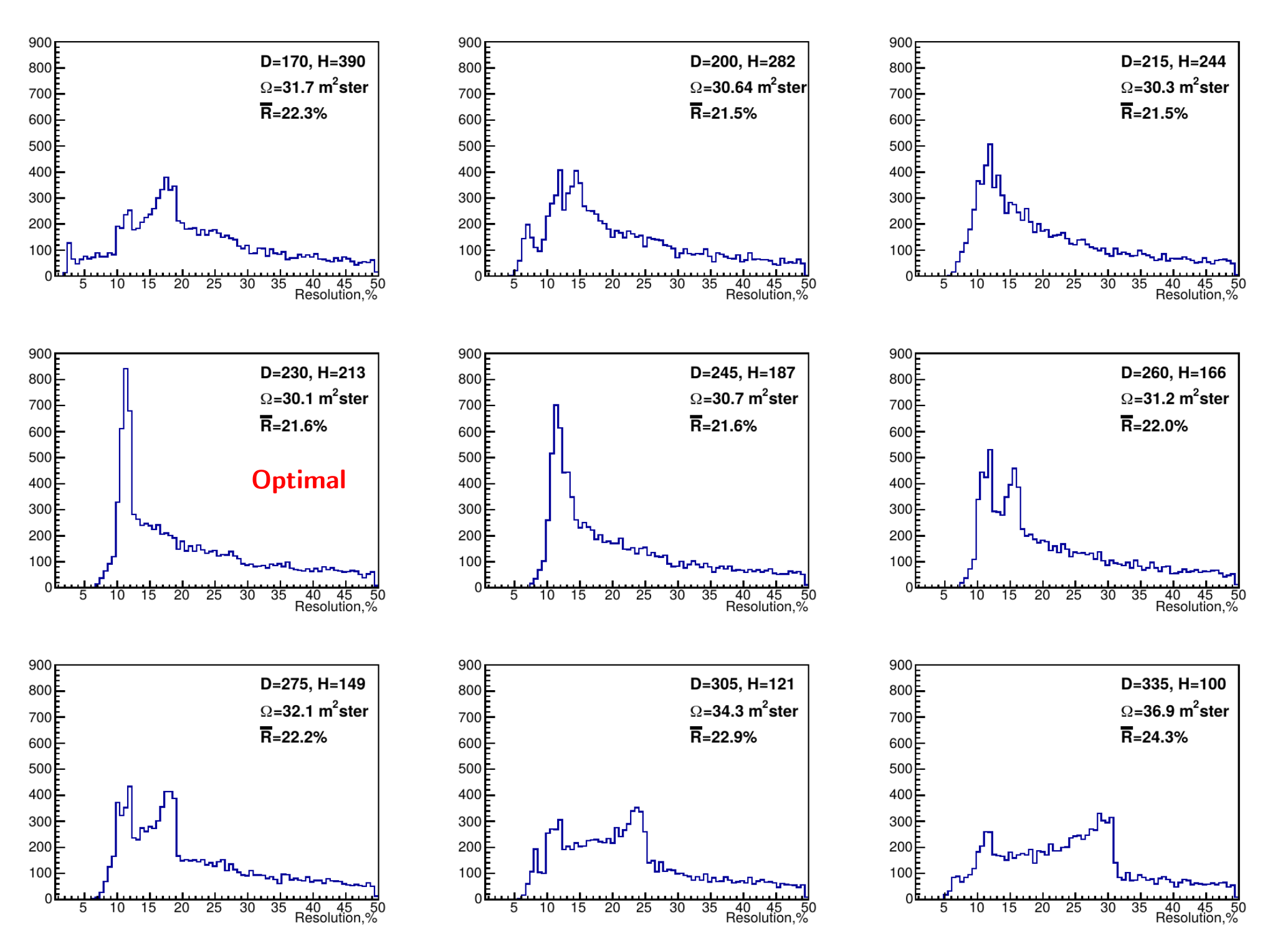}
 \caption{Energy resolution distributions for a 30 tons calorimeter and various calorimeter shapes. Notation as in \fig{Distrib10}. \label{fig:Distrib30}}
 \end{center}
\end{figure*}

\begin{figure*}[ht!]
 \begin{center}
 \includegraphics[width=0.8\textwidth]{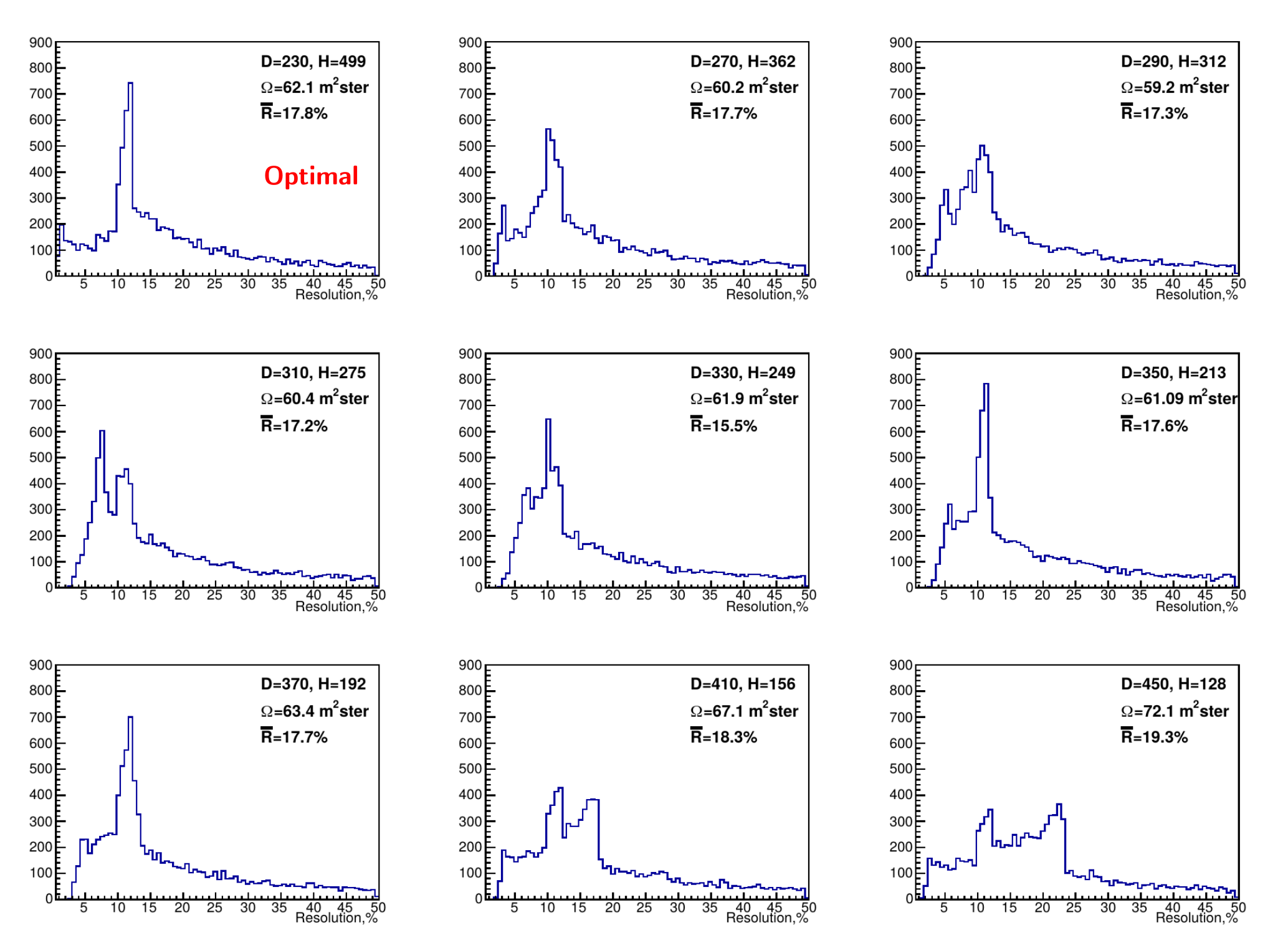}
 \caption{Energy resolution distributions for a 70 tons calorimeter and various calorimeter shapes. Notation as in \fig{Distrib10}.. \label{fig:Distrib70}}
 \end{center}
\end{figure*}

Fig.\ref{fig:Distrib10}--\ref{fig:Distrib70} show the calculated energy resolution distributions for 10, 30 and 70 tons calorimeters, respectively, and for different calorimeter shapes. The diameter of the circumscribed circle of the prism $D$ and the height of the calorimeter $H$ in the diagrams are expressed in centimeters. The graphs also show the effective geometric factor (for protons that have passed all selection conditions, including the trigger condition) and the energy resolution averaged over the distribution.

Let's pay attention to the complex shape of histograms, which is characterized by the presence of two clearly visible peaks, apart from other subtle details, which also sometimes occur. These two main peaks correspond to groups of trajectories passing either through two prism bases or through a pair of opposite lateral faces. The number of trajectories passing through these pairs of faces, with the zenith (with respect to these faces) angles cosine close to unity, is relatively large, which gives the corresponding peaks. The figures show how these peaks regularly shift as the dimensions ratio changes. Starting from a configuration with a small diameter and high height (``pencil'') as the diameter increases and height decreases (to maintain the total specified mass of the calorimeter), to a configuration with a large diameter and low height (``pancake''), the peaks regularly move towards each other, intersect at a diameter approximately equal to the height, and diverge again in different directions. With this evolution of the shape, as can be seen from the figures, the geometric factor of the device changes rather weakly, therefore, when choosing the optimal shape of the calorimeter, one should not focus on it, but pay attention to the features of the behavior of the energy resolution distribution functions.

In 10 and 30 ton calorimeter configurations, the situation is similar and the criteria for choosing the optimal shape of the calorimeter are similar as well. The most advantageous configuration corresponds to the intersection of the two peaks of the energy resolution distribution function mentioned above. In this case, they overlap and generate one peak, which is located at relatively high resolutions. That is, the most probable resolution for the calorimeter is high. For a 10 ton calorimeter the optimal dimensions are $D = 160$ cm, $H = 147$ cm, when the main peak of the distribution is located at a resolution of 18\%. The average resolution turns out to be 28\% with a geometric factor of 11.8 m$^2$sr. For a 30 ton calorimeter, the optimal dimensions are $D = 230$ cm, $H = 213$ cm, when the main peak of the distribution is located at a resolution of 11\%, and the average resolution turns out to be 22\% with geometric factor of 30.1 m$^2$sr. Such configurations are convenient in that many homogeneous events with good energy resolution are most likely to be obtained. By limiting the length of the trajectories so that the main peak is included, but trajectories with a lower resolution are not included, at the cost of a certain drop in the effective geometric factor, it will be possible to obtain spectra of very high quality, which is important for obtaining precision spectra at particle energies slightly lower than the extreme ones attainable for a calorimeter of a given mass. To obtain spectra at extremely high energies and, accordingly, with lower particle statistics, all trajectories can be included in the analysis at the cost of reducing the average energy resolution.

The situation is somewhat different for a 70 ton calorimeter (\fig{Distrib70}). Although the migration of the two main peaks described above occurs here as well, a new circumstance arises. For an extremely elongated calorimeter (``pencil''), there are quite a few trajectories that give an extremely high energy resolution -- at the level of 1–-5\%, regardless of the particle energy. Such a resolution is currently achievable only in magnetic spectrometer experiments like AMS-02 \cite{AMS-02-2015-PRL-p}, but only at energies no higher than several hundred GeV and with geometric factors that are orders of magnitude lower than the one of the HERO observatory. Since the energy resolution averaged over the entire distribution and the geometric factor are weakly dependent on the shape of the calorimeter (see \fig{Distrib70}), the the possibility of obtaining an extremely high resolution turns out to be a decisive factor when choosing a calorimeter configuration. Thus, the optimal calorimeter shape for a mass of 70 tons is a calorimeter with dimensions $D = 230$ cm, $H = 499$ cm (``pencil''), when the main peak of the distribution is located at a resolution of 11\%, the average resolution turns out to be 18\% and geometric factor is 62.1 m$^2$sr. The position of the peak is determined by the trajectories passing through the opposite lateral faces of the calorimeter, and it is the same as that of the optimal 30 tons calorimeter, but an important advantage of this configuration is a significant part of the distribution corresponding to resolutions better than 10\% and even 5\%.

\begin{figure*}[ht!]
 \begin{center}
 \includegraphics[width=0.8\textwidth]{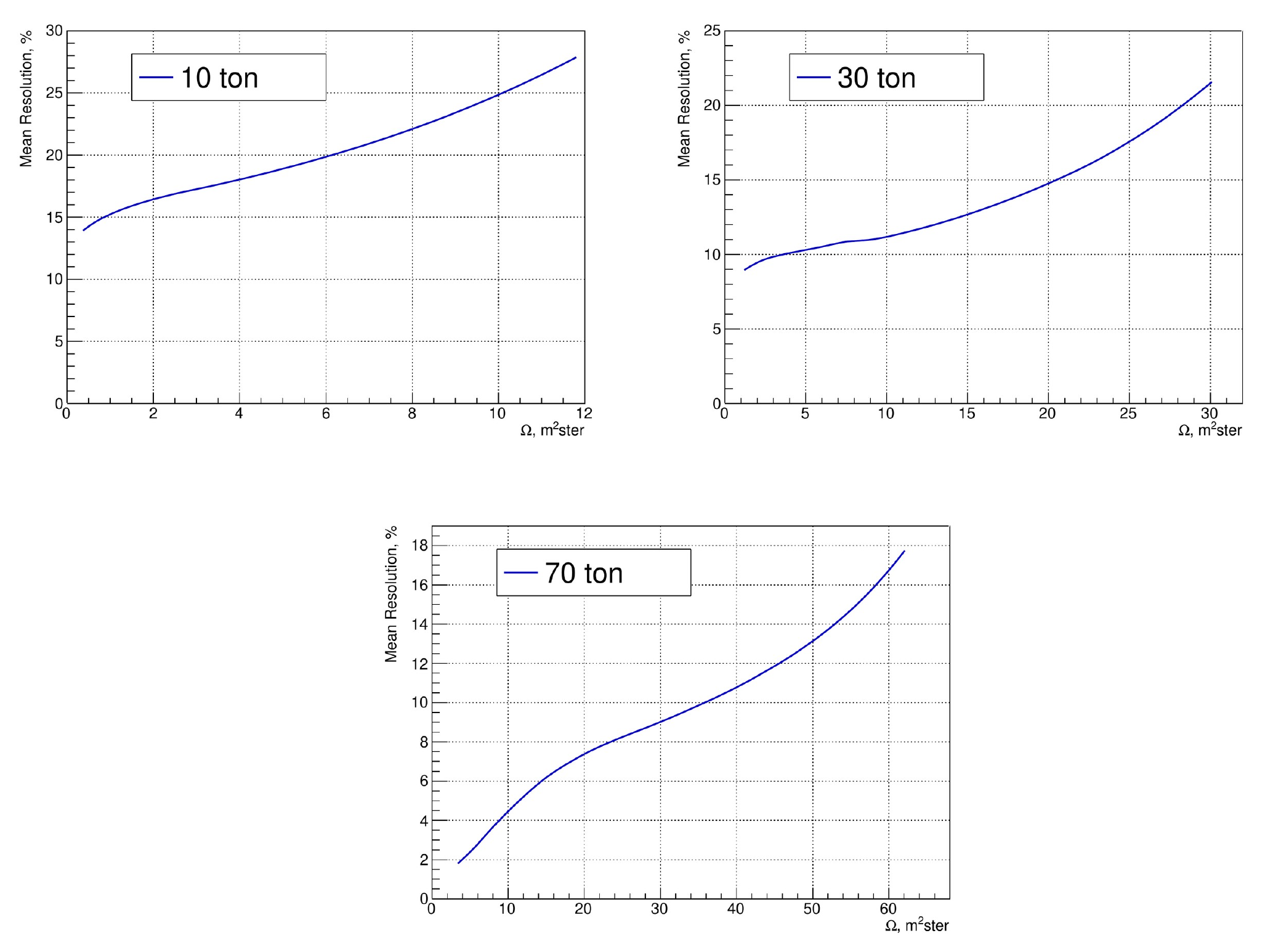}
 \caption{Dependence of the average resolution of the calorimeter on the effective geometric factor for artificial filtering of events along the length of the part of the trajectory contained in the volume of the calorimeter for calorimeters of different masses. \label{fig:OmegaR}}
 \end{center}
\end{figure*}

Working with a calorimeter of any mass and configuration, one can achieve an improvement in the average energy resolution by selecting trajectories with a sufficiently long part contained in the calorimeter volume. The cost, however, will be the loss of some of the trajectories suitable for analysis, which leads to a drop in the effective geometric factor as the resolution improves. \fig{OmegaR} shows the dependence of the calorimeter average resolution on the effective geometric factor for artificial filtering of events along the length of the part of the trajectory contained in the calorimeter volume for calorimeters of various masses and optimal configurations chosen for them (see above). It can be seen that by reducing the geometric factor for any calorimeter, the resolution can be improved, but a particularly dramatic effect occurs for a 70 tons calorimeter. For example, keeping the geometric factor 10 m$^2$sr, which is almost equal to the total maximum geometric factor of a calorimeter of 10 tons, an average energy resolution of 4.5\% can be obtained, and for the geometric factor of  4 m$^2$sr (which is still very high by the modern standards of any orbital spectrometer of cosmic rays and even greater than the whole geometric factor of the proposed HERD apparatus \cite{HERD-2019}), we obtain a resolution of 2\%, which corresponds to the energy resolution of best modern gamma-ray telescopes.

\begin{figure*}[ht!]
 \begin{center}
 \includegraphics[width=0.8\textwidth]{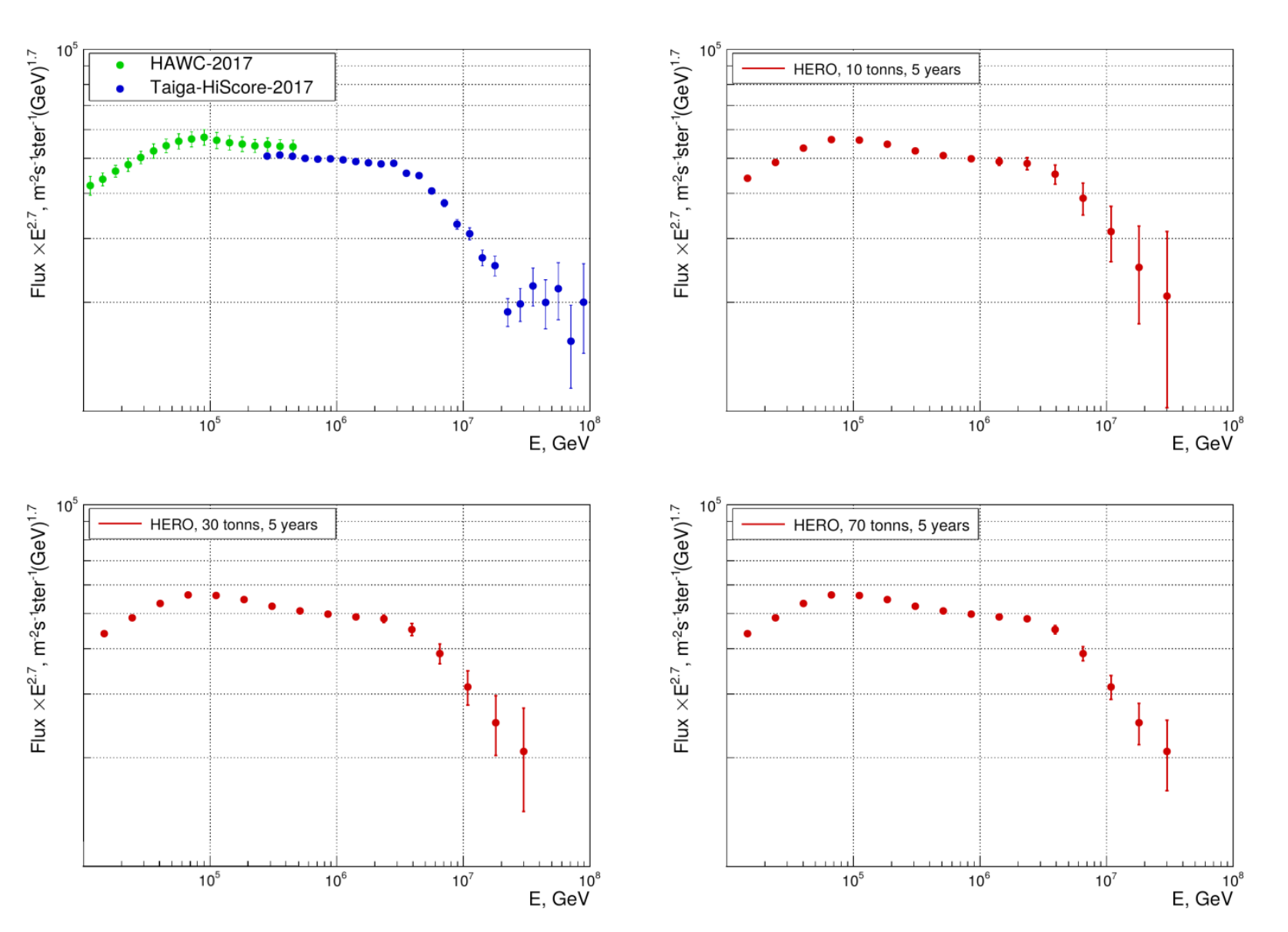}
 \caption{Expected statistics for calorimeters 10, 30 and 70 tons over 5 years of exposure for the spectrum of all particles in comparison with the statistics achieved so far in the best ground-based EAS experiments HAWC \cite{HAWC-2017} and Taiga-HiScore \cite{TUNKA-2019}. \label{fig:Statistics}}
 \end{center}
\end{figure*}

\fig{Statistics} shows the expected statistics for 10, 30, and 70 tons calorimeters over 5 years of operation in orbit when measuring the spectrum of all particles, compared with statistics achieved so far in the best ground-based EAS experiments. It should be noted that only statistical errors here are taken in account; exact expected systematic errors will be studied later. We expect that, due to the large calorimeter depth, the systematic uncertainties due to the energy containment will be no higher than 1\%.  It can be seen that in the region adjacent to the cosmic ray knee near 3 PeV from the low-energy side the HERO statistics are comparable to the results of ground-based EAS arrays, but we recall that, in contrast to EAS arrays, the HERO experiment provides element-by-element resolution in the nuclear charge. Since these statistics are very high, the HERO experiment should comprehensively describe the structure of cosmic ray spectra near the knee with elemental charge resolution. At slightly lower energies, on the order of 100 TeV, the statistics of HERO surpasses the statistics of ground-based EAS experiments, despite the fact that HERO provides an elemental analysis of the spectra. Within reasonable limitation of statistics (see \fig{OmegaR}) it will be possible to obtain precision spectra with energy resolution of a few percent in this region (especially when using a 70 ton calorimeter).

\subsection{Optimization of the strip structure of the calorimeter and reconstruction of particle trajectories}

To optimize the strip structure of the calorimeter using the FLUKA system, several mathematical models of the calorimeter were investigated. As already known from previous HERO studies, the optimal ratio of tungsten absorber and scintillator in a layered structure should be close to one radiation unit of tungsten (about 3 mm in thickness) per 20 mm of scintillator thickness. While simulating different strip systems, we proceeded from the same proportion of absorber and scintillator. Two main classes of models have been investigated. In the first class of models, which in our terminology is denoted as T2, for each absorber layer there are two scintillator layers, each of which consists of scintillation strips oriented mutually orthogonal in two adjacent planes, which provides information on the spatial organization of the shower in two mutually-orthogonal coordinates, allowing one to restore its structure. In the second class of models, denoted T3, for each layer of the absorber there are three scintillator layers, which are also organized in a strip manner, but in each successive layer the strips are rotated 60$^\mathrm{o}$ relative to each other (in the model names, the letter T is an abbreviation of the word Type, and the number denotes the number of scintillator planes per absorber layer). In the T3 model, the coordinate information in the horizontal layers is redundant.

The thickness of the scintillator strips was chosen in all cases to be equal to 8 mm simply because the scintillator sheets in the standard industrial design have such a thickness. Therefore, in the T2 model, there are 16 mm of scintillator per absorber layer, and in the T3 model there are 24 mm of scintillator. The tungsten thickness was selected according to the stated proportion (one radiation unit of absorber per 20 mm scintillator): 0.8 and 1.2 radiation units of tungsten in each layer for T2 and T3 models respectively. Obviously, in the T3 model, the calorimeter layers are thicker than in the T2 model, therefore, with the same weight and geometry of the calorimeter, the coordinate information of the T3 model is more separated in space (there are fewer absorber layers). However, in each horizontal layer in the T3 model, the coordinate information is redundant and, in this sense, is more accurate than in the T2 model. Which of these models will be more optimal can be found only by detailed simulation of both models and comparing the results.

In both of T2 and T3 model classes, calorimeters with strip pitch of 1.25 cm, 2.5 cm, and 5 cm were simulated (which covers the entire reasonable range of strip widths). It is clear that the accuracy of shower structure reconstruction depends on the strip width, which is important for reconstructing its axis, but as the strip width decreases, the number of electronic channels increases, which increases the installation cost. It is necessary to choose the optimal strip width. All simulations was carried out only for a 10 ton calorimeter, since all the features of the showers reconstruction can be clarified already for a calorimeter of only one weight: these features depend mainly on the linear dimensions of the calorimeter, and they change, roughly speaking, only as the cubic root of the weight (that is, rather slowly). Thus, a total of 6 10 ton calorimeter models were investigated, three models for both T2 and T3 classes.

Protons with initial energies of 1, 10, and 100 TeV and gamma quanta with energies of 100, 300, 1000, and 3000 GeV were studied as incident particles. The problem of shower axes reconstruction for protons is more difficult than for all heavier nuclei. Gamma quanta are important in terms of the potential for gamma astronomy with the HERO observatory. Thus, totally $6 \times 7 = 42$ separate Monte Carlo simulations were carried out with several thousand events in each simulation. Sumulations were performed for particles with isotropic incidence while taking into account the Earth's shadow, trigger condition and trajectory separation as before. The energy for protons was bounded from above by 100 TeV in this work, since even higher energies require a very large amount of calculations, which was not possible at the present stage of the work. This will be done later, but for now the parameters of interest for the models for higher energies can be obtained by extrapolating the three energy points 1, 10 and 100 TeV.

From the point of view of optimizing the strip structure of the calorimeter, the most important problem is the determination of the shower axes reconstruction accuracy. The answer to this question is important, first of all, for the primary particle charge determination algorithms (or for the confirmation of charge absence when observing gamma quanta), since the particle charge is sought as a response of the silicon matrix in the vicinity of the point of intersection of the reconstructed shower axis with the silicon charge detector. The smaller the expected corridor of errors of the shower axis in the vicinity of this point, the more reliably the charge is determined. The second aspect of the same problem is the measurement of the primary particle trajectory reconstruction angular accuracy. This is especially important for gamma astronomy, since it determines the angular resolution of the HERO observatory for gamma quanta, but it is also important for general physics of cosmic rays, since it determines the angular resolution in observing the anisotropy of the directions of arrival of the hadron CR component. Thus, the task of optimizing the strip structure is to achieve the best accuracy of reconstruction of shower axes at a reasonable cost of equipment.

From a technical point of view, when it comes to the accuracy of the shower axis reconstruction, it is important how exactly the reconstruction algorithm is arranged. To build algorithms of this type, it is important to have a clear idea of what the calorimeter events look like for particles of different types and different energies. To obtain this information, the so-called portraits of events are constructed, in which the distribution of energy deposits in the scintillators of the calorimeter is shown in conventional colors, according to which the initial trajectory of the particle must be reconstructed and the initial trajectory of the particle is shown. To develop an algorithm for the reconstruction of trajectories, several hundred portraits of events were drawn and studied for each of the 42 variants of the Monte Carlo simulation of the calorimeter.

\begin{figure*}[ht!]
 \begin{center}
 \includegraphics[width=0.8\textwidth]{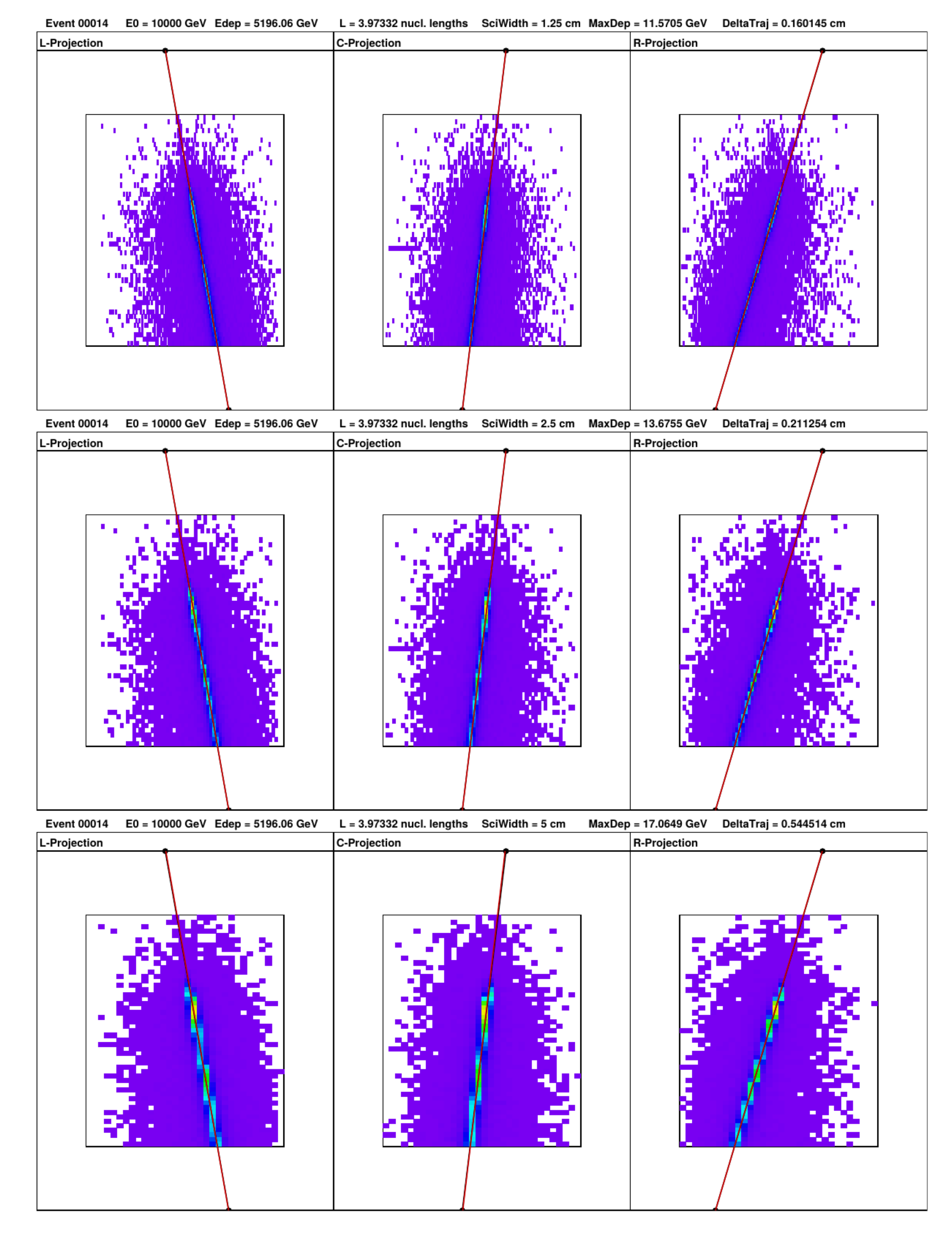}
 \caption{An example of an event portrait: a 10 TeV proton, a T3 calorimeter, from top to bottom -- different strip widths: 1.25 cm, 2.5 cm, 5.0 cm. \label{fig:Portrait-p}}
 \end{center}
\end{figure*}

\begin{figure*}[ht!]
 \begin{center}
 \includegraphics[width=0.8\textwidth]{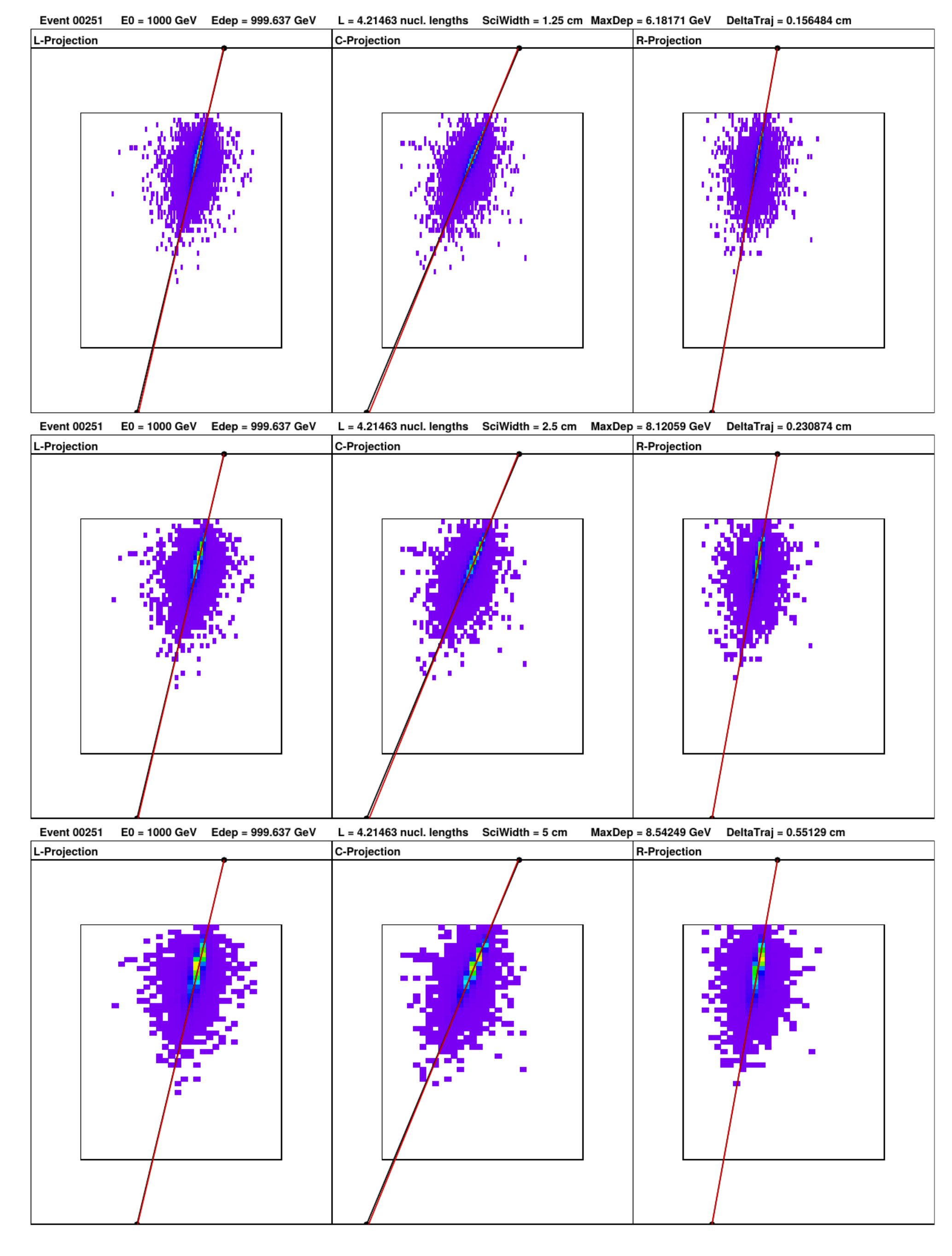}
 \caption{An example of an event portrait: 1 TeV gamma quantum, T3 calorimeter, from top to bottom -- different strip widths: 1.25 cm, 2.5 cm, 5.0 cm.\label{fig:Portrait-Gamma}}
 \end{center}
\end{figure*}

On \fig{Portrait-p} and \fig{Portrait-Gamma} typical of portraits of simulated events are shown: 10 TeV proton, T3 calorimeter; gamma quantum 1 TeV, calorimeter type T3. For each event, portraits are shown for calorimeters with different strip widths: 1.25 cm, 2.5 cm, 5.0 cm. For T3 calorimeters, all three coordinate projections of energy deposits in strips are shown in conventional colors. In addition to the pictures themselves, the title of each portrait contains some technical information about the event.

The difference in the shape of showers for protons and gamma quanta is striking: the cascades for gamma quanta are much shorter than for protons. This difference in the cascade shapes, on the one hand, makes it possible to very effectively separate hadronic CR component events from the gamma quanta and lepton (electrons and positrons, the showers shape of which are similar to the ones of gamma quanta) events. On the other hand, due to the greater length of the cascades, the accuracy of reconstruction of the shower axes for protons (and other hadrons) in the HERO calorimeter will be noticeably better than for gamma quanta and leptons.

\begin{figure*}[ht!]
 \begin{center}
 \includegraphics[width=0.8\textwidth]{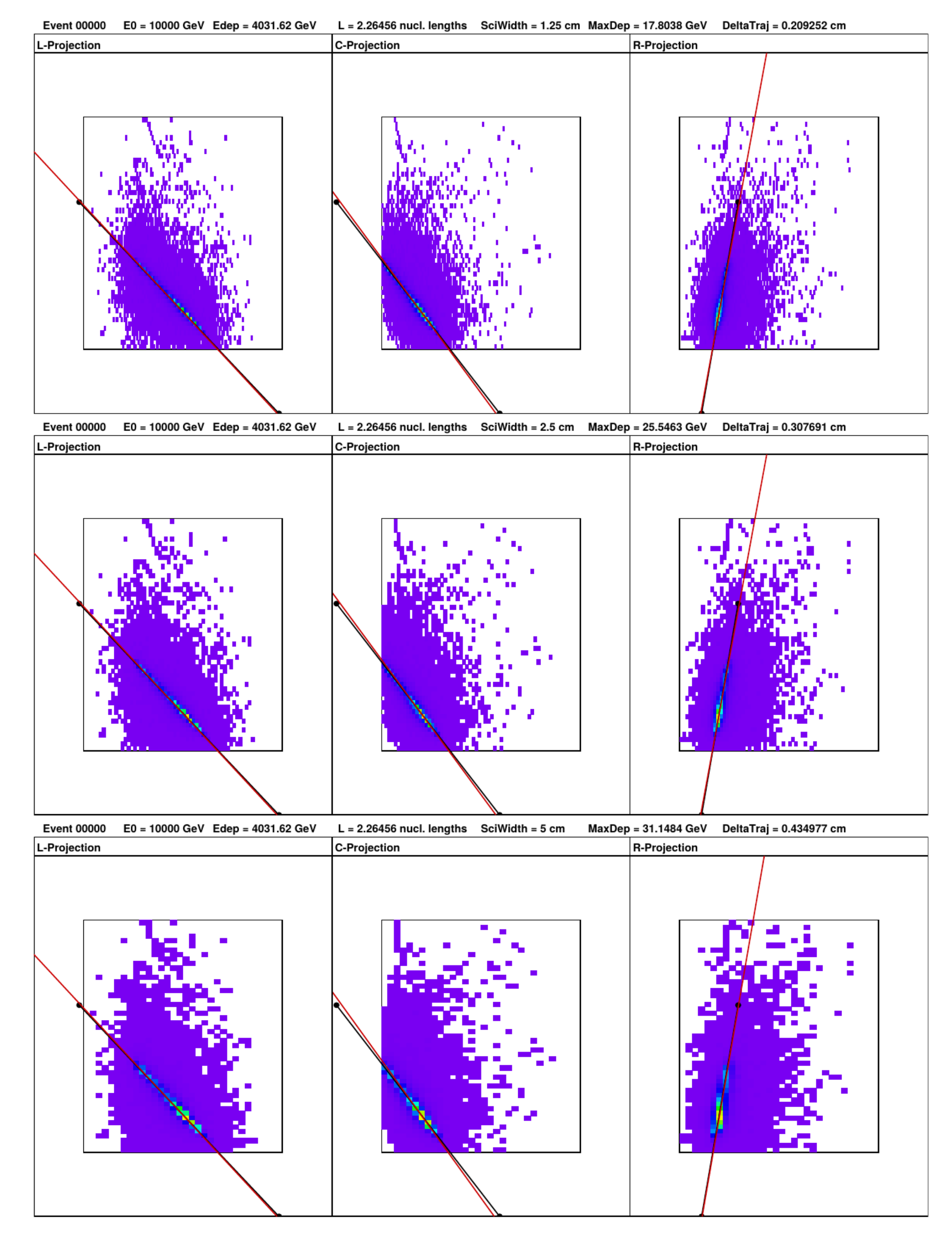}
 \caption{An example of event with a ``short'' trajectory section passing through the calorimeter: a 10 TeV proton, a T3 calorimeter, from top to bottom -- different strip widths: 1.25 cm, 2.5 cm, 5.0 cm. A noticeable discrepancy between the reconstructed and actual shower axes is seen.\label{fig:Portrait-Unprecise}}
 \end{center}
\end{figure*}

In addition to the projections of the shower axis onto the coordinate planes, which in Fig.~\ref{fig:Portrait-p} and \ref{fig:Portrait-Gamma} are shown with a black line, black circles show the projections of the point of intersection of the axis with the plane of silicon detectors, which is assumed to be located at a distance of 40 cm from the calorimeter surface and surrounds the calorimeter surface on all sides. The errors presented below in determining the position of the point of intersection of the shower axis with the plane of silicon charge detectors are determined in this geometry. The red lines show the projections of the reconstructed shower axis. Since the reconstruction accuracy is quite high for the portraits shown in Fig.~\ref{fig:Portrait-p} and \ref{fig:Portrait-Gamma}, the red line almost merges with the black one. However, this is not always the case. Fig.~\ref{fig:Portrait-p} and \ref{fig:Portrait-Gamma} show relatively simple cases of event reconstruction, when the length of the part of the particle trajectory passing through the calorimeter volume is large. \fig{Portrait-Unprecise} shows an event for a 10 TeV proton with a shorter trajectory segment in the calorimeter, and here the difference between the reconstructed and real axes is already clearly visible.

Here are the details of the developed algorithm for reconstructing the shower axes. The algorithm reconstructs the shower axis projection onto each of the coordinate planes (2 planes for the T2 configuration and three planes for the T3 configuration) and then reconstructs the entire trajectory using a set of these projections. The algorithms described below are somewhat preliminary, as they were optimized for 10 TeV protons. Further optimization is possible using the the energy deposit values in the calorimeter and the types of particles, but at this stage this problem has not yet been solved. However, it can be assumed that the improvements that can be achieved by such additional optimization will not be very significant; therefore, the algorithms described below are quite universal for solving the current problem of optimizing the calorimeter design.

Reconstruction of each projection begins with calculating the dispersions of the energy deposit distribution along the transverse coordinate of the given projection and the normalized cascade curve along the planes of this projection (i.e., along the $Z$ axis). When calculating the transverse dispersions, only strips with an energy deposit of more than 0.002 GeV are taken into account (approximately 1 MIP for scintillator plane). When there are no such strips in the plane, the dispersion is set to $-1$, which is a sign that it cannot be calculated and the contribution to the cascade curve is assumed to be zero. If the calculated dispersion turns out to be less than 0.29 of the strip width (the standard deviation for a uniform distribution of random points over the strip width), then it is artificially set to 0.29 of the strip value, since a lower value in this problem has no physical meaning.

The resulting cascade curve is used to calculate the maximum energy deposit in the plane of this projection, and then the entire cascade curve is normalized to this maximum (i.e., the maximum in the normalized cascade curve is equal to unity). Then the actual algorithm for reconstructing the projection of the axis is used, which uses the data prepared at previous stages.

The first stage of the actual axis reconstruction algorithm is the selection of the calorimeter planes for subsequent approximation of the shower axis and determination of the transverse coordinates of the shower maximum for them, which will then be used in the approximation. It is important that at this stage a number of non-trivial solutions are used, which were found by optimizing the shower axis reconstruction algorithm, namely, by minimizing the axes reconstruction errors.

\begin{figure}[ht!]
 \begin{center}
 \includegraphics[width=0.45\textwidth]{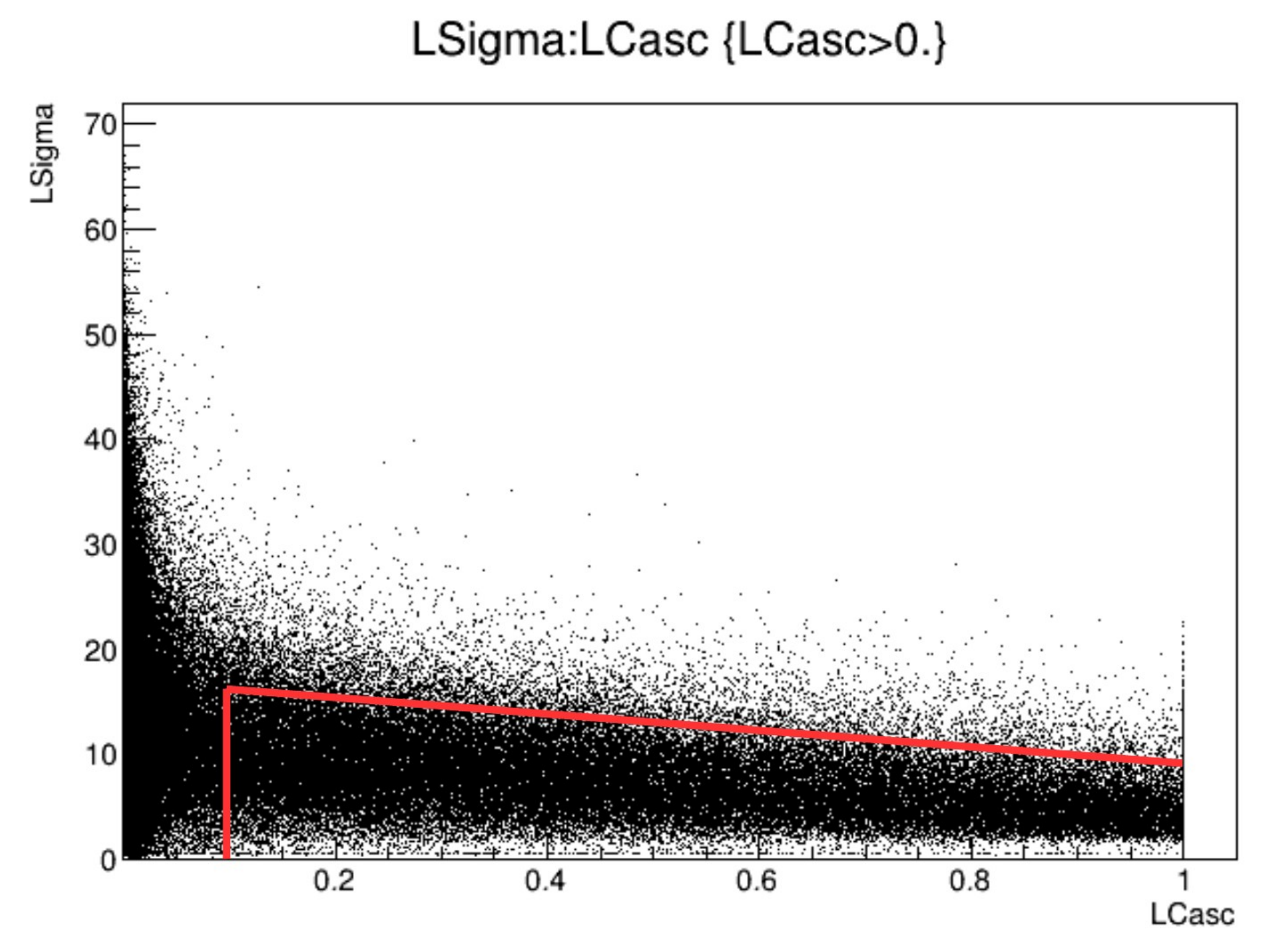}
 \caption{The admissible range of parameters on the ``relative amplitude of the cascade curve (\texttt{LCasc}) -- the value of the transverse dispersion in the calorimeter layer (\texttt{LSigma}, cm)'' plane for including the plane in the shower axis reconstruction procedure.\label{fig:LSigmaLCasc}}
 \end{center}
\end{figure}

First of all, for further work, only the calorimeter layers are selected from the admissible range of parameters on the ``relative amplitude of the cascade curve (\texttt{LCasc}) -- the value of the transverse dispersion in the calorimeter layer (\texttt{LSigma})'' plane, see \fig{LSigmaLCasc}, to include the plane in the shower axis reconstruction procedure. The area is surrounded by a red line, as can be seen in \fig{LSigmaLCasc}. This makes it possible to exclude the layers of the calorimeter, where the lateral position of the shower maximum is a priori poorly defined and can lead to a large error. As is clear from \fig{LSigmaLCasc}, this region was established based on the simulation of calorimeter events (point cloud in the same figure).

Then a preliminary estimate is made of the position of the transverse maximum in each horizontal plane. To do this, we first set the width of the smoothing pattern as 11 strips for a strip width of 1.25 cm, 5 strips for 2.5 cm, and 3 strips for 5.0 cm. Using this template, with a step of one strip along the plane of the calorimeter, the total energy deposits are sequentially searched for each sequential pattern, and among all such energy deposits, the position of the template with the maximum total energy deposit is sought (the position of the template is the position of its center). If the maximum is found at the edge of the plane, then this is considered as a sign that the shower in this plane has gone beyond the boundaries of the calorimeter, and such a point is discarded from the analysis. The found position of the maximum is a preliminary estimate of the lateral position of the shower maximum in the plane.

Then, using this preliminary position of the maximum as a center, with the template width 2 strips larger than that used to search for this preliminary maximum, the center of gravity of the energy deposit is found from the obtained template points, which is taken as the final position of the lateral shower center in this plane. Note that other ways of determining the center have been investigated, for example –– using an approximation of the energy deposit profile within the template using a Gaussian function, when there are good conditions for this (a single local maximum within the template), but the best result was obtained by simply calculating the center of gravity.

If the number of found points for approximating the axis is less than three, then the algorithm stops working, the projection of the axis is declared not found (this happens for very oblique side showers). After obtaining a set of points for the axis reconstruction, the actual approximation algorithm is launched, which performs a cycle of data approximation and correction. This cycle is structured as follows.

First, the straight line $X(Z) = a + bZ$ is reconstructed by the least squares method (LSM) based on the approximation points supplied to the input of the algorithm. In the LSM for different planes, weights are used, which are calculated as $W = \mbox{\texttt{LCasc/LSigma}}^2$. The choice of weights for the LSM is the result of several experiments with different weights. This approximation also gives the root-mean-square deviation of the transverse coordinates of the maxima in the planes from the obtained straight line $\sigma X$. Second, all deviations of the shower maxima in the planes from the obtained curve are tested, and if the deviation of the plane in absolute value exceeds three standard deviations, then the corresponding plane is excluded from the set of planes used to reconstruct the projection. After that, it is checked whether the number of planes for the reconstruction of the axis has become less than three, and when this happens, the algorithm is terminated. After discarding planes with outliers, the axis projection is reconstructed anew, and the algorithm is repeated until there are no outliers or the algorithm is aborted.

The result of the projection reconstruction algorithm are the coefficients of the straight line $X(Z) = a + bZ$, as well as the estimate of the quality of the approximation, which is calculated as the standard deviation of the maxima in the layers from the straight line, reduced to the direction perpendicular to the shower axis: $\sigma X_{Corrected} = \sigma X/\sqrt(1 + b ^ 2)$. If the projections of the trajectories in all coordinate planes are determined, then the reconstructing algorithm searches for points of intersection of the shower axis with the silicon detector and the algorithm for determining the angles of the axis is launched. First of all, this algorithm determines the average estimate of the approximation quality $\sigma X_{CorrectedMean}$ for all used planes (two for the T2 configuration and three for T3). After that, the coordinates of the shower axis intersection with the silicon detector are calculated. For the T2 configuration this problem is trivial, since this intersection point is simply the intersection point of the projections of the coordinate planes with the silicon detector surface (which itself is either a plane parallel to the plane of the calorimeter prism base, or the lateral surface of the cylinder, assuming that the lateral silicon detectors lie on the cylindrical surface). For the T3 configuration, there are three similar points for each pair of coordinate planes from among the three present in the T3 configuration. The final result is obtained by averaging over these three points. The angles of the axis in a spherical coordinate system are determined in a similar way. From these data, an error is obtained in determining the point of intersection of the shower axis with the surface of the silicon detector and the angular deviation of the reconstructed axis from the true axis in the event simulation.

\begin{figure*}[ht!]
 \begin{center}
 \includegraphics[width=0.7\textwidth]{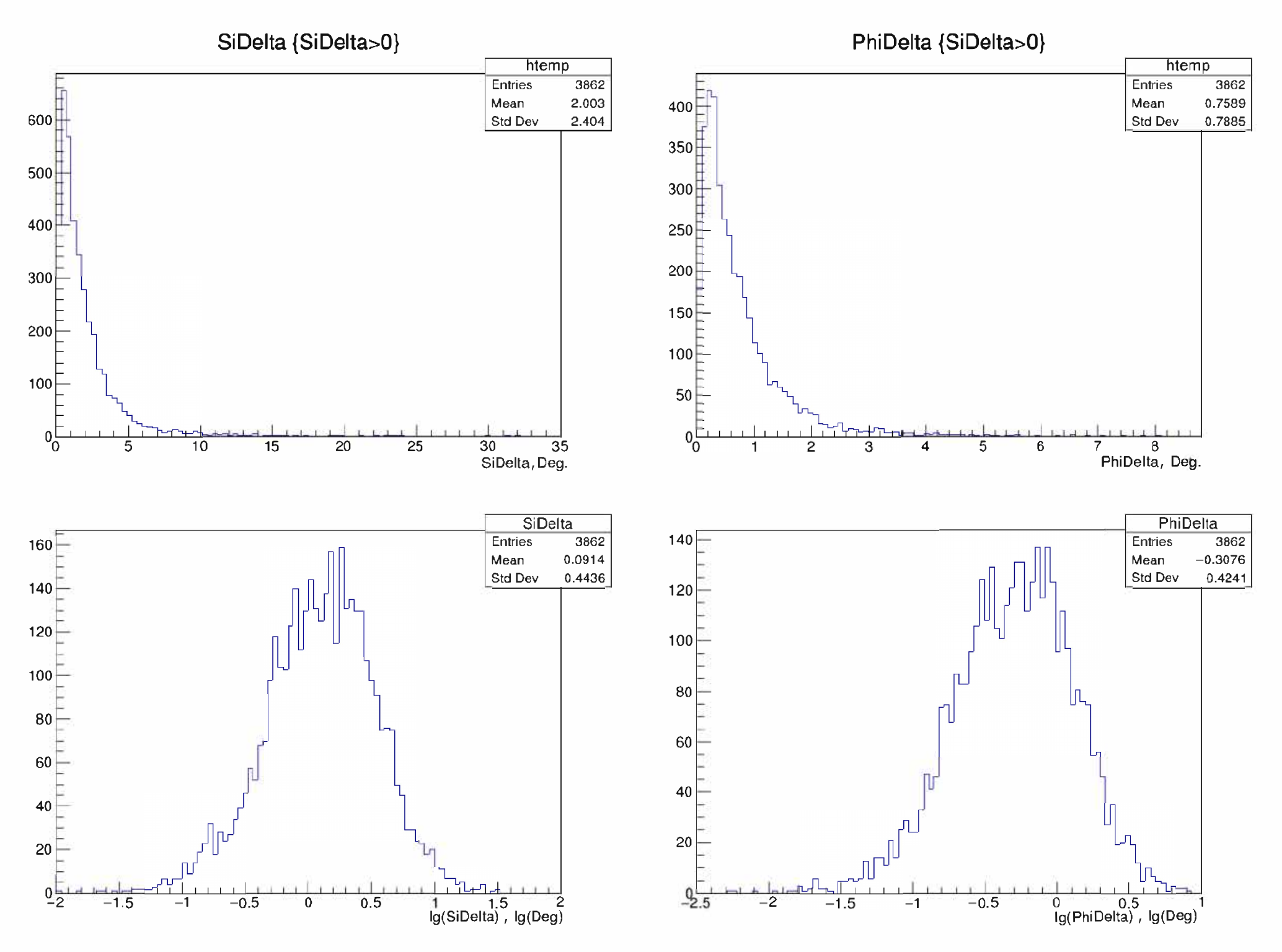}
 \caption{Examples of distributions of deviations of the reconstructed shower axis from the true axis (proton, 10 TeV, T2 calorimeter, strip width 2.5 cm). The two upper plots: the distribution of the deviations of the intersection point of the reconstructed shower axis from the true position of the intersection point, calculated along the surface of the silicon matrix (in centimeters) and the distribution of the angular error of the trajectory reconstruction (in degrees). Bottom two plots: the same as above, but the distribution for the logarithm of the corresponding errors.\label{fig:Errors-p}}
 \end{center}
\end{figure*}

\begin{figure*}[ht!]
 \begin{center}
 \includegraphics[width=0.7\textwidth]{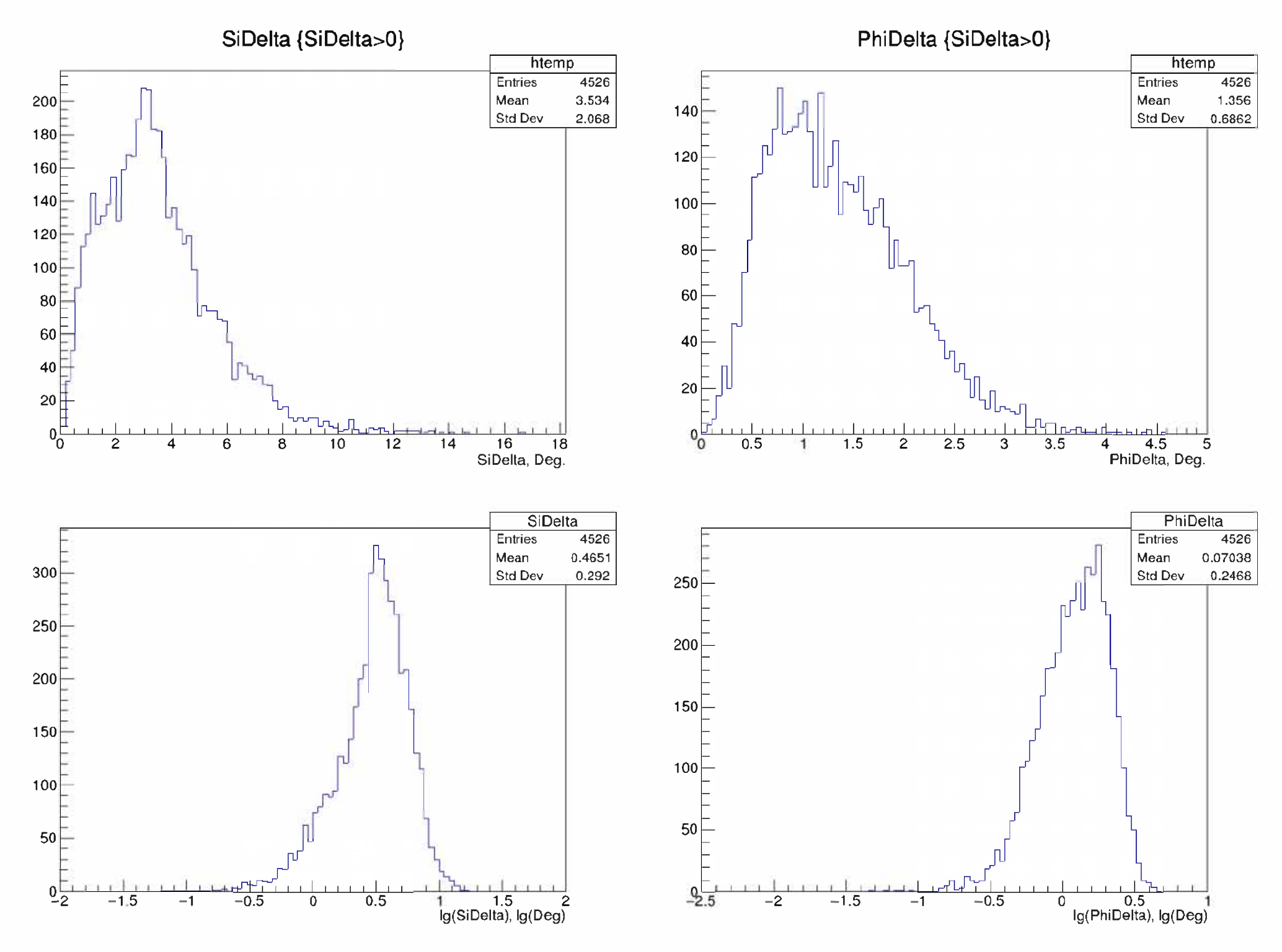}
 \caption{Examples of distributions of deviations of the reconstructed shower axis from the true axis (gamma quantum, 1 TeV, T2 calorimeter, strip width 2.5 cm). The meaning of the panels is the same as in \fig{Errors-p}.\label{fig:Errors-Gamma}}
 \end{center}
\end{figure*}

Fig.~\ref{fig:Errors-p} and \ref{fig:Errors-Gamma} show examples of the distributions of deviations of the reconstructed shower axis from the true axis. The top two plots of each figure show the distribution of the deviations of the intersection point of the reconstructed shower axis from the true position of the intersection point, calculated along the surface of the silicon matrix (in centimeters) and the distribution of the angular error of the trajectory reconstruction (in degrees). It can be seen that the distributions are highly asymmetric. However, it turns out that the distributions of the logarithms of the deviations in all cases are much more symmetric, and for protons at all studied energies they are close to normal. That is, we are dealing with distributions close to lognormal ones. The two lower plots in Fig.~\ref{fig:Errors-p} and \ref{fig:Errors-Gamma} show the distributions of the logarithms of the same parameters for which the distributions on a linear scale are shown in the upper ones. It is clearly seen that these distributions are much more symmetric. For lognormal distributions, the mean value of the logarithm is well defined, from which the logarithmic mean of the initial value can also be obtained. It is the logarithmic means of the error distributions that will be used further.

In the general case, the distribution by trajectory reconstruction errors depends on which part of the trajectory length distribution (see \fig{Distrib10}) we will use, or we will not pay attention to the lengths at all and use the entire distribution. It is technically more convenient to use a ``good'' distribution part, which starts with trajectory lengths of 3.5 nuclear lengths up to the maximum attainable values. In other words, such events include the main peak of the distribution and all events with longer trajectories. This is convenient because the bulk of the HERO results related to energies below 1000 TeV can be based on just such ``good'' events, which will be characterized by good energy resolution in the particle energy spectra. On the other hand, on longer trajectories, the axis reconstruction algorithm works more stably, therefore, according to the peculiarities of the algorithm operation in this area, it is easier to separate the operating features of the calorimeter as a whole, depending on its design. Therefore, all the data on the axis reconstruction errors given below refer to this ``good'' portion of the nuclear length distribution: \texttt{NuclLeng} $> 3.5$.

In \tab{Errors-cm} logarithmic mean errors of trajectory reconstruction in determining the point of intersection of the trajectory with the silicon matrix (in centimeters) are given; in \tab{Errors-deg}, the mean logarithmic errors of reconstruction of the trajectory angles (in degrees) are given for various calorimeter configurations, particle types, and initial energies. These tables provide a lot of material for analysis. Let us formulate some conclusions that can be drawn from the data of these tables.

\begin{table*}
 \begin{center}
 \caption{Trajectory reconstruction errors (in centimeters) in determining the point of intersection of the trajectory with the silicon matrix.\label{tab:Errors-cm}}
  \begin{tabular}{|c|c|c|c|c|c|c|}
    \hline
    \           &
    T2, 1.25 cm &
    T2, 2.5 cm  &
    T2, 5.0 cm  &
    T3, 1.25 cm &
    T3, 2.5 cm  &
    T3, 5.0 cm  \\
    \hline
    p, 1 TeV        &
    1.42 $\pm$ 0.03 &
    1.43 $\pm$ 0.03 &
    1.96 $\pm$ 0.04 &
    1.44 $\pm$ 0.04 &
    1.43 $\pm$ 0.03 &
    1.93 $\pm$ 0.04 \\
    \hline
    p, 10 TeV       &
    0.69 $\pm$ 0.02    &
    0.75 $\pm$ 0.02    &
    1.31 $\pm$ 0.03    &
    0.75 $\pm$ 0.02    &
    0.78 $\pm$ 0.02    &
    1.24 $\pm$ 0.03    \\
    \hline
    p, 100 TeV         &
    0.54 $\pm$ 0.02    &
    0.61 $\pm$ 0.02    &
    1.11 $\pm$ 0.03    &
    0.55 $\pm$ 0.02    &
    0.61 $\pm$ 0.02    &
    1.06 $\pm$ 0.03    \\
    \hline
    $\gamma$, 100 GeV  &
    2.62 $\pm$ 0.06    &
    3.16 $\pm$ 0.06    &
    5.57 $\pm$ 0.09    &
    2.76 $\pm$ 0.07    &
    3.25 $\pm$ 0.07    &
    5.13 $\pm$ 0.09    \\
    \hline
    $\gamma$, 300 GeV  &
    2.43 $\pm$ 0.06    &
    2.94 $\pm$ 0.06    &
    5.00 $\pm$ 0.08    &
    2.58 $\pm$ 0.06    &
    2.87 $\pm$ 0.06    &
    4.61 $\pm$ 0.08    \\
    \hline
    $\gamma$, 1000 GeV &
    2.31 $\pm$ 0.05    &
    2.75 $\pm$ 0.05    &
    4.67 $\pm$ 0.08    &
    2.33 $\pm$ 0.06    &
    2.62 $\pm$ 0.05    &
    4.32 $\pm$ 0.07    \\
    \hline
    $\gamma$, 3000 GeV &
    2.16 $\pm$ 0.05    &
    2.52 $\pm$ 0.04    &
    4.39 $\pm$ 0.06    &
    2.25 $\pm$ 0.05    &
    2.51 $\pm$ 0.05    &
    3.95 $\pm$ 0.06    \\
    \hline
  \end{tabular}
 \end{center}
\end{table*}

\begin{table*}
 \begin{center}
 \caption{Trajectory direction reconstruction errors (in degrees).\label{tab:Errors-deg}}
  \begin{tabular}{|c|c|c|c|c|c|c|}
    \hline
    \            &
    T2, 1.25 cm  &
    T2, 2.5 cm   &
    T2, 5.0 cm   &
    T3, 1.25 cm  &
    T3, 2.5 cm   &
    T3, 5.0 cm   \\
    \hline
    p, 1 TeV      &
    0.56 $\pm$ 0.01  &
    0.56 $\pm$ 0.01  &
    0.78 $\pm$ 0.02  &
    0.57 $\pm$ 0.01  &
    0.56 $\pm$ 0.01  &
    0.75 $\pm$ 0.02  \\
    \hline
    p, 10 TeV     &
    0.27 $\pm$ 0.01  &
    0.29 $\pm$ 0.01  &
    0.52 $\pm$ 0.01  &
    0.29 $\pm$ 0.01  &
    0.30 $\pm$ 0.01  &
    0.49 $\pm$ 0.01  \\
    \hline
    p, 100 TeV    &
    0.21 $\pm$ 0.01  &
    0.24 $\pm$ 0.01  &
    0.44 $\pm$ 0.01  &
    0.22 $\pm$ 0.01  &
    0.24 $\pm$ 0.01  &
    0.43 $\pm$ 0.01  \\
    \hline
    $\gamma$, 100 GeV    &
    1.11 $\pm$ 0.02  &
    1.35 $\pm$ 0.02  &
    2.43 $\pm$ 0.03  &
    1.17 $\pm$ 0.02  &
    1.38 $\pm$ 0.02  &
    2.20 $\pm$ 0.03  \\
    \hline
    $\gamma$, 300 GeV    &
    1.01 $\pm$ 0.02  &
    1.23 $\pm$ 0.02  &
    2.13 $\pm$ 0.03  &
    1.07 $\pm$ 0.02  &
    1.21 $\pm$ 0.02  &
    1.97 $\pm$ 0.03  \\
    \hline
    $\gamma$, 1000 GeV   &
    0.92 $\pm$ 0.02  &
    1.10 $\pm$ 0.02  &
    1.94 $\pm$ 0.03  &
    0.94 $\pm$ 0.02  &
    1.07 $\pm$ 0.02  &
    1.80 $\pm$ 0.02  \\
    \hline
    $\gamma$, 3000 GeV   &
    0.86 $\pm$ 0.02  &
    1.02 $\pm$ 0.01  &
    1.80 $\pm$ 0.02  &
    0.92 $\pm$ 0.02  &
    1.03 $\pm$ 0.02  &
    1.65 $\pm$ 0.02  \\
    \hline
  \end{tabular}
 \end{center}
\end{table*}

1. If we follow the dependence of the magnitude of the errors on the width of the scintillation strips, we can see that the smallest error occurs for the narrowest strips. With an increase in the strip width, it grows, but with a change in the width from 1.25 cm to 2.5 cm, the error increases relatively slightly, and going from 2.5 cm to 5.0 cm is much stronger. For example, for protons in the T2 configuration, averaging over all the energies used from 1.25 cm to 2.5 cm gives an increase in the reconstruction error of the point of intersection of the shower axis with the silicon detector by 7.5\%, and from 2.5 to 5.0 cm, by 64.5\%. Due to the angular error, the situation is qualitatively the same; it is the same for the T3 calorimeter and for gamma quanta in all calorimeter configurations. This means that the strip width of 2.5 cm still does not roughen the coordinate distribution of energy in the shower too much, but the width of 5 cm is already comparable to the shower width; therefore, the information becomes much more coarse and the reconstruction error greatly increases. Thus, the strip width can be taken equal to 2.5 cm, which slightly reduces the quality of the spatial resolution of the shower structure and the quality of the trajectory reconstruction, but it allows to save twice on the number of electronic channels and the power consumption in comparison with the strip width of 1.25 cm. The strip width, which is close to 2.5 cm, looks like optimal.

2. When comparing the configurations of the calorimeter T2 and T3, it is possible to restrict ourselves only to configurations with the optimal strip width of 2.5 cm selected above. From Tab.~\ref{tab:Errors-cm} and \ref{tab:Errors-deg} then it can be seen that both by the error of reconstruction of the point of intersection of the shower axis with the silicon matrix, and by the error of reconstruction of the angles, the T2 and T3 configurations give results practically not differing within the statistical errors of the simulation (errors are also given in the tables). Therefore, the configuration T2 or T3 can be selected for reasons of design simplicity. With the same strip width of 2.5 cm for the T3 configuration, the electronic channels are slightly smaller and all planes have the same design so T3 looks preferable.

\begin{figure*}[ht!]
 \begin{center}
 \includegraphics[width=0.9\textwidth]{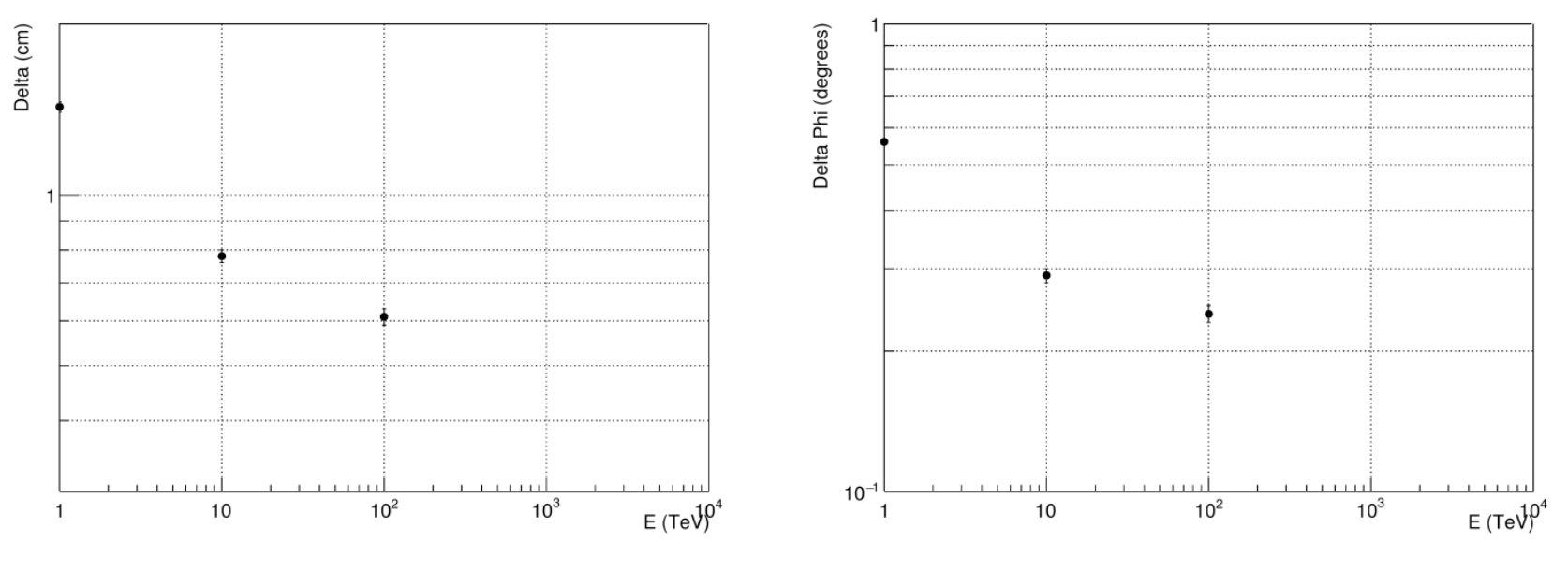}
 \caption{Left panel -- the dependence of the error in reconstructing the point of intersection of the shower axis with the silicon matrix on the proton energy in the T2 configuration for a strip width of 2.5 cm; right panel -- the error in reconstructing the axis direction under the same conditions.\label{fig:Delta-E}}
 \end{center}
\end{figure*}

3. It can be seen from Tab.~\ref{tab:Errors-cm} and \ref{tab:Errors-deg} that as the particle energy increases, the shower axis reconstruction errors tend to decrease in all calorimeter configurations and for all types of particles. \fig{Delta-E} on the left shows the dependence of the error in reconstructing the point of intersection of the shower axis with the silicon matrix on energy, on the right -- a similar dependence for the error in reconstructing the axis direction (for a proton, in the T2 configuration, 2.5 cm). Based on the behavior of the points in the left graph, one can expect that in the proton energy range between 1 PeV and 10 PeV, the average accuracy of the axis reconstruction at the level of the silicon matrix will be about 0.5 cm.This is an important value, since the size of the silicon pixels of the matrix is assumed to be 1 cm and the average axis reconstruction error turns out to be less than the pixel size. At the highest particle energies, the most important problem is the choice of the correct pixel against the background of an intense flux of particles scattered in the backward direction, and the magnitude of the expected error in the reconstruction of the axis at a level of 0.5 cm suggests that the problem of choosing a pixel with the signal of the primary particle against the background of reverse currents will be solved. The right graph of \fig{Delta-E} shows that the angular resolution of the device at the highest energies for protons will be about 0.2 degrees. However, it should be noted that the above extrapolations are not very reliable; therefore, an explicit simulation of the device operation at energies above 100 TeV is still required.

\begin{figure*}[ht!]
 \begin{center}
 \includegraphics[width=0.9\textwidth]{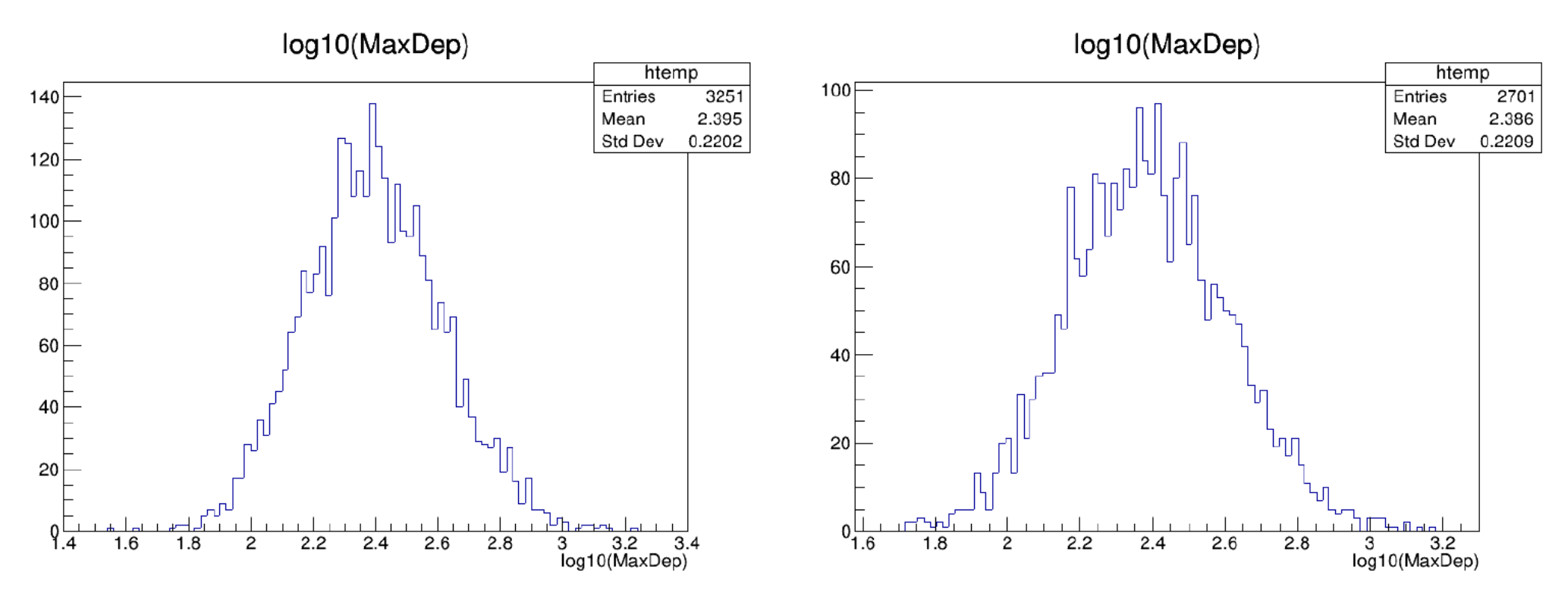}
 \caption{Distribution of logarithms of maximum strip energy deposits (in GeV) for a strip width of 2.5 cm and configurations T2 and T3 of the calorimeter, for protons with an initial energy of 100 TeV.\label{fig:Dynamic1}}
 \end{center}
\end{figure*}

The performed simulation allows solving one more important problem -- the problem of determining the required dynamic range for the measuring channels of the calorimeter scintillators. The upper limit for measuring the energy deposit in scintillators should be sufficient to measure the strongest signals that can appear there during the development of nuclear-electromagnetic showers. \fig{Dynamic1} shows the distribution of the logarithms of the maximum strip energy deposits (in GeV) for a strip width of 2.5 cm and configurations T2 and T3 of the calorimeter, for protons with an initial energy of 100 TeV. The position of the right tail of the distributions is important here. It can be seen that for the T2 and T3 configurations it is practically the same and amounts to about 1.6 TeV. Similar data can be used to obtain points for other simulated proton energies -- 1 TeV and 10 TeV, and the obtained dependencies can be extrapolated to the region of the highest energies of incident particles, for which data are supposed to be obtained.

\begin{figure}[ht!]
 \begin{center}
 \includegraphics[width=0.45\textwidth]{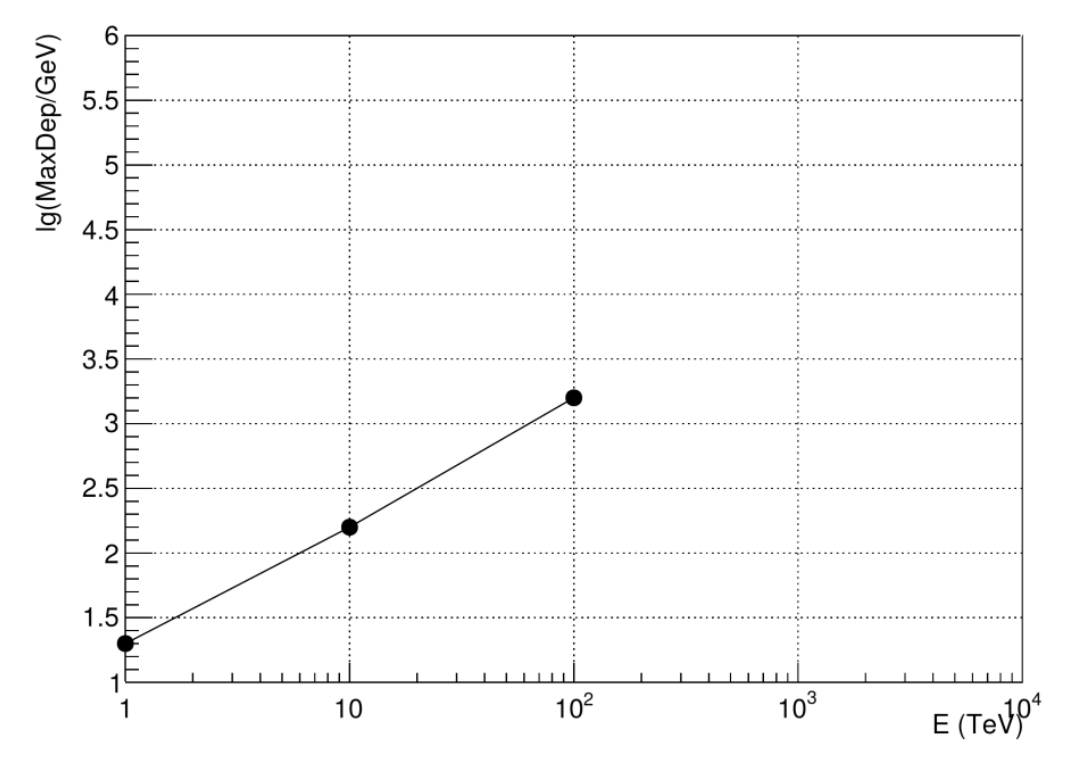}
 \caption{Dependence of the maximum energy deposit in the strip as a function of the proton energy (the strip width is 2.5 cm; for the T2 and T3 calorimeters, the dependencies are the same).\label{fig:Dynamic2}}
 \end{center}
\end{figure}

\fig{Dynamic2} shows the dependence of the maximum energy deposit in the scintillator strip as a function of the initial energy of the incident protons (strip width 2.5 cm; for calorimeters T2 and T3, the dependencies are the same). The points in this figure are based on 4800 simulated events for each one of them. It can be seen that the maximum energy deposit in the strip is practically proportional to the initial energy of the particle; therefore, it is possible to estimate the maximum energy deposit at the highest planned particle energy for the HERO experiment, that is, near 10 PeV. Extrapolation gives $\sim$160 TeV. For a completely safe operation of the calorimeter at the highest energies, this value must be increased by approximately one and a half times, which gives an admissible upper limit of the energy deposit before the appearance of nonlinearities in the measuring channel of the calorimeter of approximately 250 TeV per scintillation strip.

\subsection{Back currents}
\label{sec:backcur}

\begin{figure}[ht]
 \begin{center}
 \includegraphics[width=0.45\textwidth]{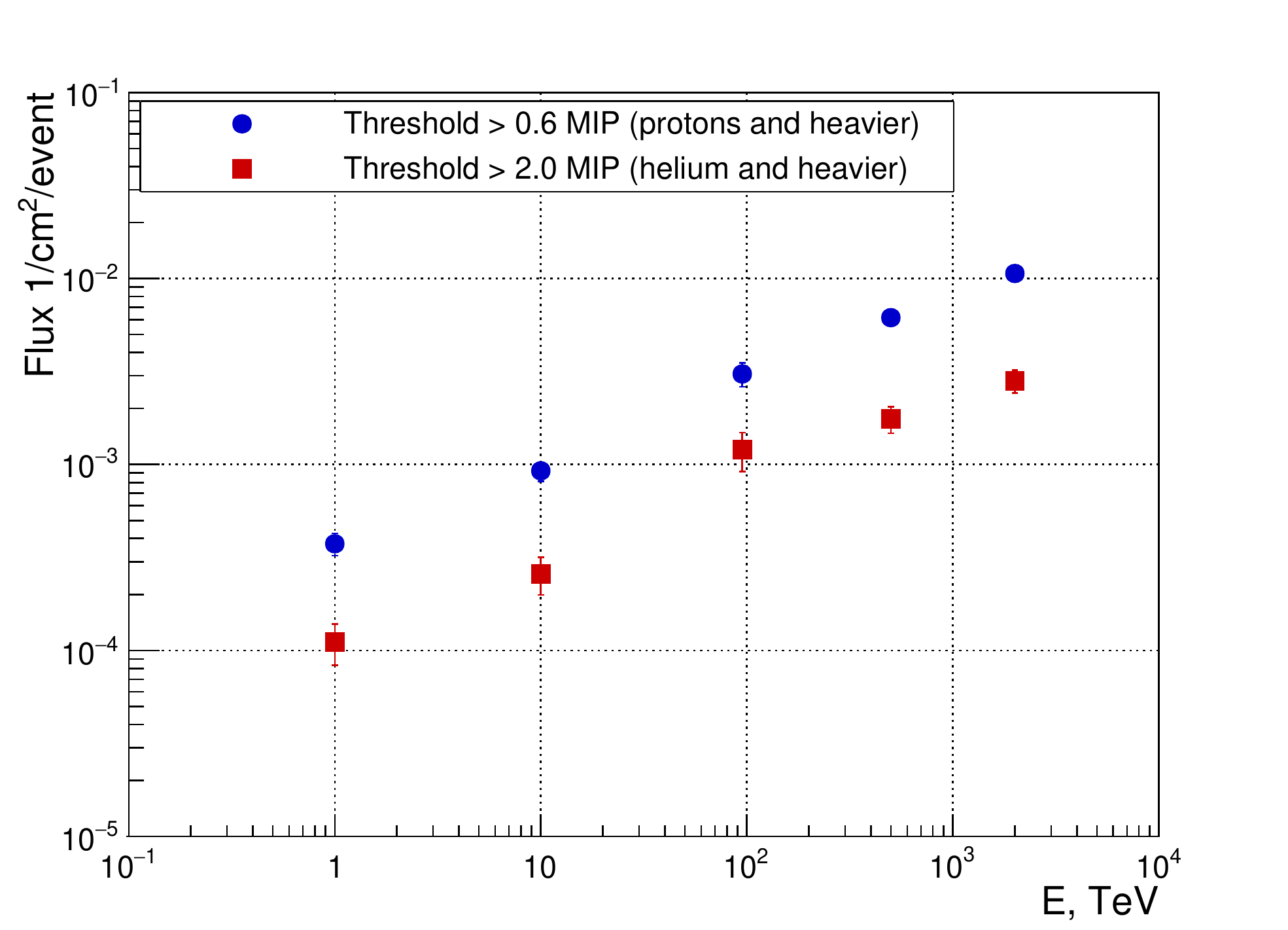}
 \caption{Probability of back current emulation of cosmic ray nuclei (in terms of back current partiles flux through $1\times1$\,cm$^2$ pixel per one event) for a proton hitting the calorimeter straight down for different proton energies and different backcurrent particles energy deposit threshold in the charge measurement silicon detector plane. \label{fig:BackCurrent}}
 \end{center}
\end{figure}

The probability of particles emulating other charges through back currents was determined with a Monte-Carlo simulation using the FLUKA package \cite{FLUKA2014}. The calorimeter with a layered structure described earlier was hit with protons at 1, 10, 95, 500 and 2000 TeV energies perpendicular to the upper face (straight down). A silicon pad detector with pixel size of $1\times1$\,cm$^2$ was placed 40 cm above the calorimeter. The trajectory of the proton (collinear to the shower axis) was intersecting the silicon pad detector exactly at the center of one of the pixels. For all pixels different from this one and for which the distance from the axis is no more than 11 cm, a mean backcurrent particles count with an energy deposit higher than a certain set threshold was determined. This count was then averaged over the 11cm diameter circle on the silicon detector. The dependence of this averaged count on the particle energy for two thresholds - 0.6\,MIP and 2\,MIP is shown on fig. \fig{BackCurrent}. A threshold of 0.6\,MIP corresponds to an emulated proton or heavier nuclei particle, and 2\,MIP corresponds to helium and heavier particles. Even for 2 PeV protons the probability of registration of an emulated back-current proton or heavier nuclei is around 1\%. For heavier nuclei this probability is around 0.3\%, and for lower initial particles energies this probability is even lower.

\section{Conclusion. About the implementation of the HERO project}

The above optimization of the calorimeter geometry makes it possible to estimate the number of electronic channels and energy consumption of the HERO observatory. Considering that the surface area of the equipment changes more slowly than its mass, with a sevenfold weighting of the device in comparison with the ten-tons version presented in the article \cite{HERO-2019}, the total number of registration channels in the system increases by about 3.5 times (or 2 times for the thirty-tons version) -- proportional to the mass to the power of 2/3. This circumstance makes it possible to preserve the main technical solutions in the readout electronics of the charge detector (CD) and ionization calorimeter (IC) and assess their feasibility. It was previously assumed that the design of the HERO would use electronics developed for the NUCLEON project, and this conclusion remains valid for all instrument configurations discussed in this article. A more accurate estimate of the number of registration channels in the charge detector and the HERO calorimeter for the three proposed weight variants of the apparatus is presented in the \tab{Channels}.

\begin{table}
 \begin{center}
 \caption{Number of electronic channels depending on the weight of the calorimeter\label{tab:Channels}}
  \begin{tabular}{|c|c|c|c|}
    \hline
    Weight  &
    10 t    &
    30 t    &
    70 t    \\
    \hline
    CD channels &
    270 000     &
    570 000     &
    1.050 000   \\
    \hline
    IC channels &
    6 700       &
    14 000      &
    32 700      \\
    \hline
  \end{tabular}
 \end{center}
\end{table}

The power consumption of the device in the first approximation is proportional to the number of registration channels, thus, based on the assessment of the power consumption of the ten-tons version made in \cite{HERO-2019} and the estimate of the number of channels for different overall-mass options given in \tab{Channels}, we can estimate the power consumption of the three HERO variants as follows:
\begin{itemize}
 \item 10 tons -- power consumption near 5    kW
 \item 30 tons -- power consumption near 10.5 kW
 \item 70 tons -- power consumption near 19.5 kW
\end{itemize}

The daily scientific data volume of HERO changes at first approximation according to the geometric factor. Considering that it was previously proposed to implement a two-threshold data collection mode -- with a low energy threshold (less than 300 GeV) and a high data collection rate and with a high energy threshold (above 1000 GeV) and a low data collection rate, it is possible to estimate the daily volume of scientific information of the apparatus as shown in \tab{Data}.

\begin{table}
 \begin{center}
 \caption{Daily volume of data depending on the weight of the calorimeter\label{tab:Data}}
  \begin{tabular}{|c|c|c|c|}
    \hline
    Weight           &
    Low threshold    &
    High thershol    \\
    \hline
    10 t    &
    800 GB/day &
    100 GB/day \\
    \hline
    30 t    &
    2400 GB/day &
    300 GB/day \\
    \hline
    70 t    &
    4800 GB/day &
    600 GB/day \\
    \hline
  \end{tabular}
 \end{center}
\end{table}

It can be seen from the above estimates that with a high energy threshold, the daily volume of scientific data for the heaviest variant of the apparatus does not exceed the maximum data volume for the lightest variant. This makes it possible to use a unified system for collecting and transmitting scientific information for all three HERO variants (as well as the same ground-based data reception complex) -- as in the least loaded ten-tons version, compensating for the increase in loads in heavier versions by a temporary increase in the energy registration threshold.

The specificity of the space experiment HERO determines the main requirement for the space complex to be able to take out the maximum mass outside the atmosphere. Therefore, the main requirement for the HERO space complex is the requirement for the launch vehicle.

Currently, Russia uses heavy-class missiles "Proton-M" (payload in low reference orbit is 23.7 tons) and "Angara-A5" (24.5 tons). Several new types of heavy and super-heavy launch vehicles with the ability to launch into low reference orbit up to 125 tons of payload expected being prepared for commissioning by 2029, so the planned launch date is no earlier than that. We hope that HERO observatory could be used as one of the first payloads in test flights of super-heavy launch vehicles.

% \section{Acknowledgments}
% Acknowledgments should be inserted at the end of the paper,
% before the references, not as a footnote to the title. Use the
% unnumbered Acknowledgements Head style for the Acknowledgments
% heading.

% %% Bibliography
% %% Author year style
% \bibliographystyle{plain}
% % \bibliographystyle{model5-names}
% %\bibliographystyle{plain}
% %\biboptions{authoryear}
% \bibliography{Knee}

\end{document}